\documentclass[a4paper,11pt]{article}
\pdfoutput=1
\usepackage{jinstpub} 
\usepackage{longtable}
\usepackage{lscape}   
\usepackage{upgreek}
\usepackage[utf8]{inputenc}
\usepackage{dcolumn}
\usepackage{bm}
\usepackage{multirow}
\usepackage{xcolor}
\usepackage{lineno}

\begin{document}

\title{
Application of zone refining to the development of NaI(Tl) detectors for 
SABRE North}

\newcommand{\UniMI}{%
  Dipartimento di Fisica ``Aldo Pontremoli'', Universit\`a degli studi di Milano,\\
  Via G. Celoria, 16, 20139 Milano, Italy%
}
\newcommand{\INFNMI}{%
  Sezione INFN di Milano, Via G. Celoria, 16, 20139 Milano, Italy%
}
\newcommand{\Unisalento}{%
  Dipartimento di Matematica e Fisica ``Ennio De Giorgi'', Universit\`a del Salento,\\
  Via Arnesano, 73100 Lecce, Italy%
}
\newcommand{\INFNLE}{%
  Sezione INFN di Lecce, Via Arnesano, 73100 Lecce, Italy%
}
\newcommand{\UniRM}{%
  Dipartimento di Fisica, Universit\`a di Roma ``La Sapienza'',\\
  Piazzale Aldo Moro, 2, 00185 Roma, Italy%
}
\newcommand{\UniPD}{%
  Dipartimento di Fisica e Astronomia ``Galileo Galilei'', Universit\`a degli Studi di Padova,\\
  Via F. Marzolo, 8, 35131 Padova, Italy%
}
\newcommand{\INFNRM}{%
  Sezione INFN di Roma, Piazzale Aldo Moro, 2, 00185 Roma, Italy%
}
\newcommand{\LNGS}{%
  INFN Laboratori Nazionali del Gran Sasso,\\
  Via Giovanni Acitelli, 22, 67100 Assergi (AQ), Italy%
}
\newcommand{\Mellen}{%
  The Mellen Company Inc, \\
  40 Chenell Dr.~Concord, NH, U.S.A.%
}
\newcommand{\RMD}{%
  Radiation Monitoring Devices Inc.\\
  Watertown, Massachusetts, 02472, U.S.A.%
}
\newcommand{\Jiagellonian}{%
  Jagiellonian University, Faculty of Physics, Astronomy and Applied Computer Science, \\
  M. Smoluchowski Institute of Physics, Lojasiewicza~11, 30--348 Krakow, Poland%
}

\newcommand{\Canfranc}{%
  Laboratorio Subterr\'aneo de Canfranc, \\
 Paseo de los Ayerbe S/N, 22880, Canfranc-Estaci\'on, Spain%
}


\author[a,b]{C.~Ananna,}
\author[c,d]{F.B.~Armani,}
\author[b]{G.~Cataldi,}
\author[c,d]{D.~D'Angelo,}
\author[e]{G.~D'Imperio,}
\author[a,b]{M.L.~De Giorgi,}
\author[f]{G.~Di Carlo,}
\author[f]{A.~Di Giacinto,}
\author[g]{M.~Diemoz,}
\author[f,1]{A.~Ianni
\note{Corresponding author: aldo.ianni@lngs.infn.it},}
\author[a,b]{S.G.~Khattak,}
\author[d]{E.~Martinenghi,}
\author[b]{A.~Miccoli,}
\author[h]{M.~Misiaszek,}
\author[a,b,2]{D.~Montanino
\note{Corresponding author: daniele.montanino@unisalento.it},}
\author[e]{V.~Pettinacci,}
\author[f]{L.~Pietrofaccia,}
\author[e,g]{S.~Rahatlou,}
\author[f]{K.~Szczepaniec,}
\author[e]{C.~Tomei,}
\author[c,d]{V.~Toso,}
\author[f]{C.~Vignoli,}
\author[b,i]{S.~Zuhra} 
\affiliation[a]{\Unisalento}
\affiliation[b]{\INFNLE}
\affiliation[c]{\UniMI}
\affiliation[d]{\INFNMI}
\affiliation[e]{\INFNRM}
\affiliation[f]{\LNGS}
\affiliation[g]{\UniRM}
\affiliation[h]{\Jiagellonian}
\affiliation[i]{\UniPD}

\author[j]{\emph{(SABRE North Collaboration)}, \\ L.~Cid,}
\author[k]{A.~Mellen-Spencer,}
\author[f]{S.~Nisi}
\author[l]{and J.~Tower}

\affiliation[j]{\Canfranc}
\affiliation[k]{\Mellen}
\affiliation[l]{\RMD}

\abstract{The SABRE North experiment is developing ultra-high radiopurity NaI(Tl) detectors to investigate dark matter. To achieve this, SABRE North utilizes the technique called \emph{zone refining} for NaI purification. This work details the mathematical model developed to describe the purification process. By comparing this model to the results of the commissioning and production runs conducted prior to crystal growth, the distribution coefficients were determined for various impurities, contained in the powder at the parts-per-billion (ppb) level. Furthermore, the synthesis of data from both zone refining and normal freezing is discussed. These findings can be used to predict the SABRE North detectors  background level in the energy region-of-interest for dark matter search and to optimize the production of ultra-high purity crystals through multiple purification strategies.
}
\keywords{Materials for solid-state detectors; Dark Matter detectors (WIMPs, axions, etc.); Scintillators, scintillation and light emission processes (solid, gas and liquid scintillators); Detector design and construction technologies and materials}
\maketitle

\section{Introduction}

The SABRE North detector is foreseen for deployment at the Gran Sasso National Laboratory (LNGS) in Italy, where it will search for the annual modulation signal induced by dark matter particle interactions \cite{drukier1986}. 
The SABRE North setup will feature at least nine ultra-high purity NaI(Tl) detectors, each weighing approximately 5~kg, housed inside a passive shielding of copper and polyethylene. 
The experiment aims to achieve a background rate in the 1$\div$6~keV region-of-interest (ROI) less than 1~counts/day/kg/keV (dru). To date, such a background level has only been achieved by the DAMA/LIBRA experiment, where potassium—a dominant background source—was maintained at a level of $\leq 20$~ppb after powder purification and crystal growth \cite{bernabei2008, bernabei2020}. 

Achieving this goal requires extreme care in the selection and purification of the NaI powder and strict control during crystal growth. 
The strategy adopted by SABRE North relies on: (1) the selection of powder with a reduced presence of (radio)contaminants with respect to commercial powder, (2) its purification via zone refining \cite{suerfu2021}, and (3) the growth of crystals with the Bridgman-Stockbarger method \cite{bridgman1925, bridgman, suerfu2020}.

The primary background sources in the ROI have been identified as $^{40}$K, $^{3}$H, $^{210}$Pb, $^{87}$Rb, and the $^{238}$U, $^{232}$Th chains that may be present in the crystal bulk \cite{bernabei2008, calaprice2021}. 
In standard commercial NaI powder, these contaminants are typically present at ppm levels or higher. A powder with a reduced content of these elements, later named ``Astro Grade'' was originally developed by Sigma-Aldrich in collaboration with the SABRE project \cite{shields2015, sabre_north, sabre_south} (Astro Grade was later supplied by Merck) and is currently the starting product for the SABRE-North crystals. On average, Astro Grade powder contains potassium at $<$20~ppb, lead at the 1~ppb level, and uranium and thorium at $<$0.1~ppb. An alternative NaI powder with similar radiopurity has also been developed as part of the COSINE200 project \cite{park2020, keon2023}.

Current studies suggest that Astro Grade purity is insufficient to reach the SABRE North target background \cite{calaprice2021,antonello2021, calaprice2022}.
For this reason, SABRE-North has developed a zone refining technique as a critical step to further purify the powder prior to crystal growth. \newline\indent
In this work, we present the first application of the zone refining technique to reach ppb level of contaminants at the scale necessary to produce about 50~kg of crystals—the minimum mass required for the proposed physics case~\cite{nai47}. We demonstrate that zone refining is highly effective in removing  critical contaminants, most notably $^{40}$K. 

This work is structured as follows. In Sec. \ref{SecII} the assay method used to determine contaminant concentrations in the NaI powder, in zone refining runs, as well as  after crystal growth, is discussed. Sec. \ref{SecIII} address the concept of zone refining, with reference to the mathematical model that has been developed for the purpose of exploiting this technique in the framework of SABRE North. The experimental procedure adopted is described in Sec. \ref{SecIV}. Results and data analysis are reported and discussed in Sec. \ref{SecV}. In Sec. \ref{SecVI} the findings of this work are summarized briefly and comparison with other methods and literature are commented. In Sec. \ref{SecVII} conclusions and comments on perspectives are reported.

\section{Contaminants and assay methods}\label{SecII}

In low-counting experiments such as SABRE North, the radio-purity assay of detector components is a crucial prerequisite, that can be performed using various techniques. For the purposes of the present study, inductively coupled plasma mass spectrometry (ICP-MS) was adopted~\cite{Heusser1995,Laubenstein2020}. This method  measures the concentration of contaminants in small quantities of sample material, allowing both rapid screening and sensitivity at the sub-ppb level. Within the framework of SABRE North, ICP-MS was used to determine contaminant levels in the starting NaI powder, in the material that underwent purification by zone refining, and in ingot samples following the crystal growth. The semi-quantitative analysis method was used, for which it is standard practice to assume an uncorrelated systematic uncertainty of around 20-30\% \cite{nisi_ref, nisi_ref1} on each measurement. In this work, a 20\% systematic uncertainty has been assumed. ICP-MS measurements reported in this work were performed by Seastar Chemicals. Complementary measurements were performed at LNGS and at the  Laboratorio Subterr\'{a}neo de Canfranc (Spain).
For the purpose of SABRE North, ICP-MS provides concentration measurements for many elements, including $^{238}$U (with 99.27\% isotopic abundance), $^{232}$Th (with $\sim$100\% isotopic abundance), $^{85}$Rb (with 72.2\% isotopic abundance), $^{39}$K (with 93.3\% isotopic abundance), and $^{208}$Pb (with 52.4\% isotopic abundance),
$^{40}$K (0.0117\% natural abundance), $^{87}$Rb (27.8\% natural abundance), and $^{210}$Pb through beta decay produce a crucial contribution to the background in the ROI. In particular, $^{40}$K decays by electron capture (10.7\% branching ratio) with the emission of  a 3~keV X-ray just in the middle of the ROI. Therefore, it is crucial to reduce this contaminant as much as possible prior to growth.
$^{238}$U and $^{232}$Th chains produce a number of beta decays which contribute to the total rate in the ROI. However, at the contamination level detected in the Astro Grade this contribution is marginal. A special remark is in order for $^{210}$Pb, which has a beta decay half-life of 22.2 years and it is produced in the last part of the $^{238}$U chain. An activity of $^{210}$Pb at the level of 0.5-1~mBq/kg in the crystal, as detected in the SABRE project \cite{calaprice2021,nai47}, corresponds to a contamination of $1.8-3.6\times 10^{-19}$~g/g, which is far below any detection limit by ICP-MS. Therefore, while for $^{40}$K and $^{87}$Rb we can infer the contamination through the ICP-MS measurement of $^{39}$K and $^{85}$Rb, for $^{210}$Pb, in order to quantify the effectiveness of the purification method, we need to assume that all physical processes involved in zone refining and crystal growth equally affect both $^{208}\mathrm{Pb}$ and $^{210}\mathrm{Pb}$. Under this assumption, we trace $^{208}$Pb to understand the possible behavior of  $^{210}$Pb.

It is worth noting that a direct measurement of the radioactive activity of $^{40}\mathrm{K}$, $^{87}\mathrm{Rb}$, and $^{210}\mathrm{Pb}$ at the levels required by SABRE North is only feasible in underground conditions, where the low cosmic rays flux enhances the sensitivity to rare processes, and requires a long exposure, of at least several months.

\section{Zone refining}\label{SecIII}

High-purity crystalline materials are essential for advanced technologies, including radiation detection and optical systems, where trace impurities can significantly degrade performance \cite{zhang2018}. This is particularly critical in searches for rare events, such as dark matter interactions or neutrinoless double beta decay ($0\nu\beta\beta$), which generally require impurity levels at the sub-ppb scale for radioactive contaminants. Zone refining is a primary technique for achieving these stringent purity requirements. First introduced by Pfann \cite{pfann1952, pfannbook}, it is a directional solidification method widely employed to purify scintillating crystals, such as sodium iodide (NaI), and semiconductors, such as silicon (Si) and germanium (Ge). It is extensively utilized in the production of high purity germanium detectors (like for $0\nu\beta\beta$ experiment LEGEND-200 \cite{legend200,Gradwohl:2020kms}). 
Unlike chemical purification methods, zone refining minimizes contamination risks as it requires no solvents or additional reactants. In zone refining the final purity obtainable is dictated by the initial powder contamination.

\begin{figure}[t]
 	\centering
 	\includegraphics[width=0.8\textwidth]{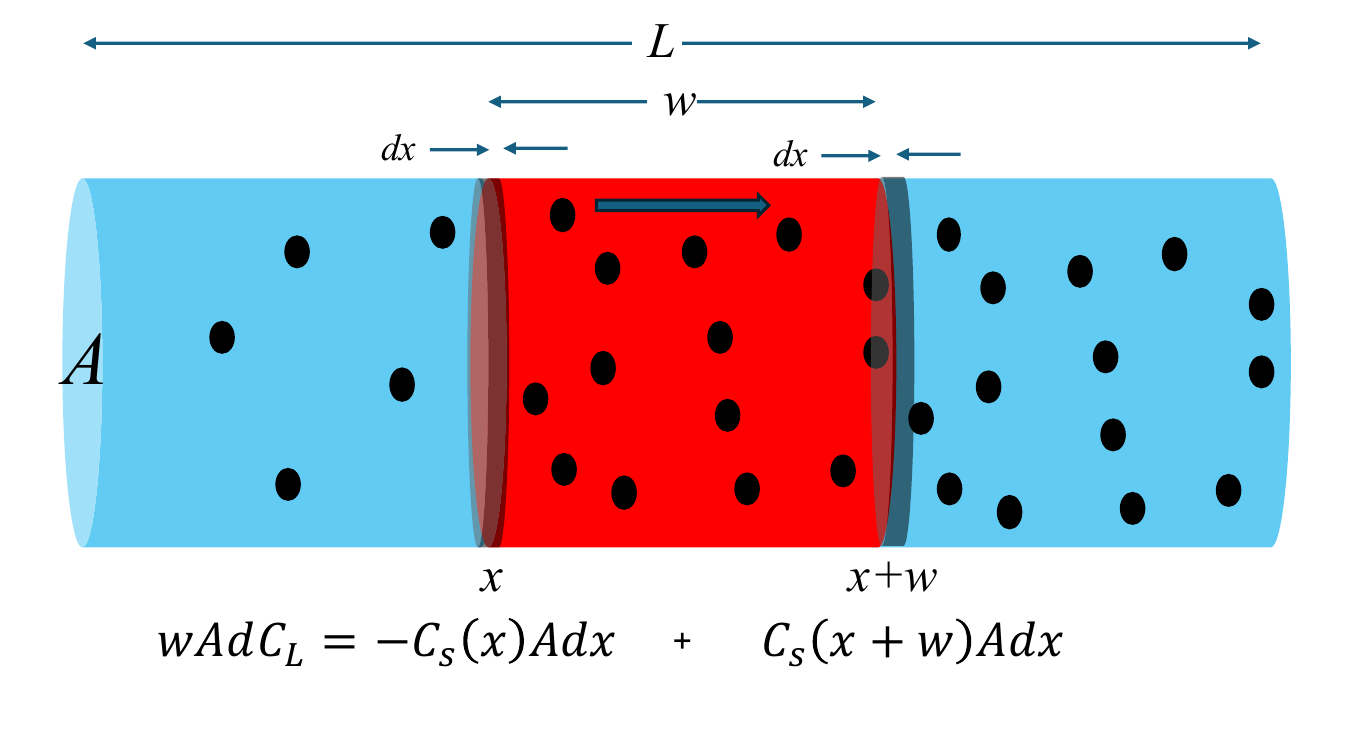}\vspace{-0.5cm}
 	\caption{Working principle of zone refining. The blue zones are solid while the red zone is liquid. The dots show the contaminant concentration in the different zones. $C_L$ and $C_S$ are the concentrations in the liquid and solid, respectively. The molten zone is moving from left to right.}
 	\label{fig:ZR_figure}
 \end{figure}

The technique exploits the different solubility of impurities in the molten and solid phases, governed by the  distribution (or segregation) coefficient, $k$. This coefficient is defined as the ratio of the contaminant concentration in the solid phase ($C_S$) to that in the liquid phase ($C_L$), $k = C_S / C_L$, at the solid-liquid interface \cite{pfann1952,spim2000b}. For most contaminants, $k$ is less than one. During the process, a heating coil creates a localized molten zone along a crystalline rod (the charge). As the heating coil advances, contaminants with $k<1$ concentrate in the melt due to their higher mobility in the liquid phase. Through repeated passes, impurities are progressively transported to one end of the charge for subsequent removal, leaving behind a highly purified bulk material \cite{spim2000b}. In Fig.~\ref{fig:ZR_figure} we show the working principle of zone refining: the charge length is $L$, the molten zone width is $w$ and the molten zone is moving from left to right.

The efficiency of zone refining depends on several interdependent experimental parameters that require careful optimization. We assume that the heating coil moves very slowly in such a way that the contaminant is always uniformly distributed in the molten zone. An effective distribution coefficient $k_{\rm eff}$ is adopted to describe the contaminant segregation \cite{BCS}:
\begin{equation}
k_{\rm eff}=\frac{k}{k+(1-k)\exp\left(-\frac{V\delta}{D}\right)}\,\, ,\label{eq:keff}
\end{equation}
where $V$ is the moving speed of the heating coil, $D$ the diffusivity of the contaminant in the melt and $\delta$ is the thickness of the diffusion boundary, that is the stagnant layer of fluid adjacent to a surface where molecular diffusion takes place. The parameters $D$ and $\delta$ are unknown, but we expect $D\sim 10^{-4}\div 10^{-5}$~cm$^2/$s and $\delta\lesssim 1$~mm \cite{Tahara,Kim}. Moreover, the Marangoni convection can affect the thickness of the boundary layer impacting on the refining efficiency \cite{wang2014}. With a velocity $V\lesssim 10^{-4}$ cm/s $D/\delta \gg V$ and $k_{\rm eff} \simeq k$. The zone refining for SABRE North takes place at a speed of order $10^{-5}$ cm/s ensuring this condition is met.

The zone travel speed is a critical factor that balances refining efficiency and process duration: slower speeds enhance impurity removal by optimizing $k_{\text{eff}}$ (reducing it for $k < 1$ or increasing it for $k > 1$) 
but prolong the processing time. Additionally, the molten zone length, which is determined by the heating power, travel speed, and thermal properties of the crucible and charge, 
impacts both impurity kinetics and energy consumption. The temperature gradient at the solid-liquid interface further dictates micro-segregation patterns and defect formation. Finally, the number of zone passes must be optimized; while additional passes improve purity by concentrating contaminants at the terminal regions, they also increase time and energy costs \cite{cheung2008, rodway1989}. 

While traditional batch methods are often limited by time inefficiency, continuous or uninterrupted zone refining addresses these issues through specialized designs for sustained operation. To optimize such systems, researchers utilize theoretical and mathematical models to refine parameters like temperature gradients and 
travel speeds, aiming to enhance scalability while balancing energy constraints \cite{spim2000b, cheung2008b}. For the SABRE North experiment, we have adopted an uninterrupted zone refining protocol and developed a specialized mathematical model to optimize the yield and efficiency of this purification method.

It's worth mentioning that zone refining in certain compounds can alter the stoichiometric ratio and is therefore not suitable to subsequent crystal growth using purified material. This is not the case for NaI, however. Consequently, when using it, the different behavior of the constituents in the compound with regard to the segregation effect must be taken into account \cite{Jackson,Swider}.

\subsection{Fundamental principles and mathematical models}

Assuming a cylindrical ingot of total length $L$ with a  cross section $A$ (see Fig.~\ref{fig:ZR_figure}), defining  $w$ ($w<L$) as the width of the molten zone (that is related to the width of the heating coil), we can write the balance equation for the variation of the solute in the molten zone when the heating coil moves from the position $x$ to $x+dx$:
\begin{equation}
wA dC_L(x)=-C_S(x)Adx+C_S(x+w)Adx\,\, .\label{eq:balance}
\end{equation}
the first term is the variation of the solute in the molten zone, the second the quantity of solute entering in the ``freezing zone'' and the third the quantity of solute entering in the molten zone. Eq.~(\ref{eq:balance}) is valid only if the heating coil does not overcome the end of the ingot, $x+w\leq L$. Using $C_S = k C_L$ we write the equation
\begin{equation}
\frac{dC_{S,n}(x)}{dx}=\frac{k}{w}\left[C_{S,n-1}(x+w)-C_{S,n}(x)\right]\,\, ,\label{eq:ZR1}
\end{equation}
where the labels $n$ refer to the concentration of solute after $n$ passages of the heating coil. For each passage, the starting condition is determined imposing that a fraction $k$ of the average concentration of the molten zone enters the tip of the ingot ($x=0$)
\begin{equation}
C_{S,n}(0)=k\int_0^w C_{S,n-1}(x)\,dx\,\, ,\label{eq:start}
\end{equation}
When the heating coil overcomes the end of the ingot, the balance equation becomes
\begin{equation}
A d\left[(L-w)C_L(x)\right]=-C_S(x)Adx\,\, ,\label{eq:balance1}
\end{equation}
which has an analytic solution 
\begin{equation}
C_S(x)=C_S(L-w)\left(\frac{L-x}{w}\right)^{k-1}\,\,\,\, {\rm for}\,\,\,\, L-w<x\leq L\,\, .\label{eq:tail}
\end{equation}
The concentration is thus always divergent in the tail of the ingot if $k<1$ due to the fact that the pollutant is moved toward the tail of it. Hereafter, for simplicity we will drop the subscript $S$.
For an initial constant concentration $C(x)=C_0$ the solution after the first passage is
\begin{equation}
C_1(x)=C_0\left[1-(1-k)e^{-\frac{kx}{w}}\right]\,\,\,\, {\rm for}\,\,\,\, 0<x\leq L-w\,\, .\label{eq:balance2}
\end{equation}
However, Eq.~(\ref{eq:ZR1}) together with Eqs.~(\ref{eq:start}) must be integrated numerically to obtain the concentration after $n$ passages. 

 \begin{figure}[b!]
 	\centering
 	\includegraphics[width=0.9\textwidth]{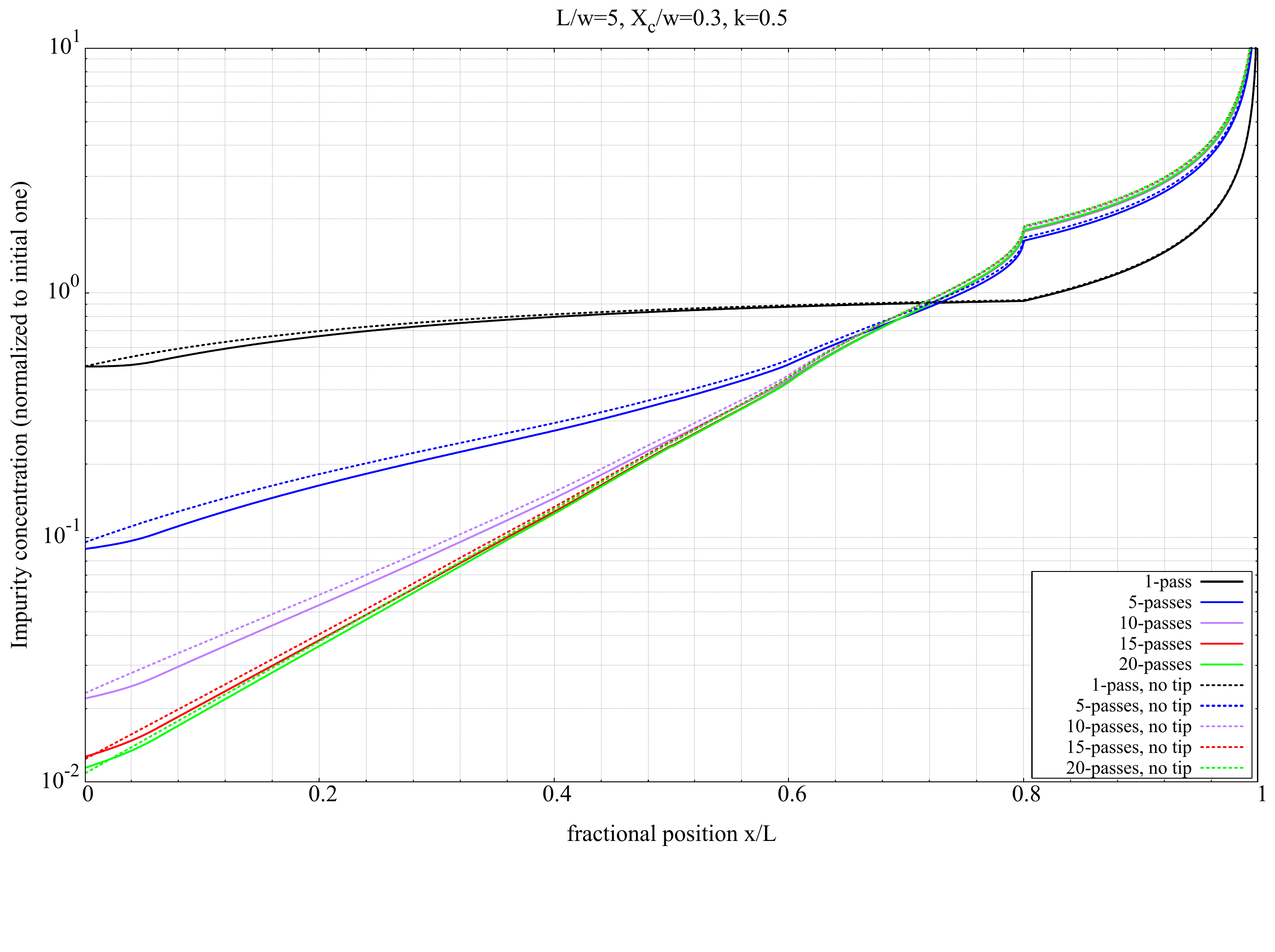}\vspace{-1cm}
 	\caption{Distribution of impurities for $k=0.5$ and $L/w=5$ as function of position $x$ normalized to ingot length $w$ for 1, 5, 10, 15, 20 passes. The effect of including a cone tip with $X_C/w=0.3$ are shown (dashed lines)}
 	\label{fig:distribution_k05_short_ampoule}
 \end{figure}

For $n\to\infty$ Eq.~(\ref{eq:ZR1}) tends to a limiting (``ultimate'') equation
\begin{equation}
\frac{dC_\infty(x)}{dx}=\frac{k}{w}\left[C_\infty(x+w)-C_\infty(x)\right]\,\, .\label{eq:ZRlim}
\end{equation}
This equation can be solved by an exponential distribution $C_\infty(x)=\alpha e^{\beta x}$ with $\beta$ solution of the transcendent equation $\beta=k(e^\beta-1)$. The coefficient $\alpha$ can be found imposing that the total quantity of solute is conserved after all passages
\begin{equation}
\alpha=\frac{\beta L/w}{e^{\beta L/w}-1}\,\, .\label{eq:alpha}
\end{equation}
However, the exponential solution for the ultimate distribution is only approximate because for the tail the true solution is the power law in Eq.~(\ref{eq:tail}) discussed below. 

We notice that the equations are independent of the constant cross section from the ingot $A$. If the cross section is variable $A=A(x)$, Eq.~(\ref{eq:ZR1}) can be generalized to the following equation
\begin{equation}
C'_n(x)V(x)=k[C_{n-1}(x+w)A(x+w)-C_n(x)A(x)]-C_n(x)V'(x)\,\, ,\label{eq:ZR2}
\end{equation}
where
\begin{equation}
V(x)=\int_x^{x+w} A(x')dx'
\end{equation}
is the volume of the molten zone, and the starting condition becomes
\begin{equation}
C_n(0)=\frac{1}{V(0)}\int_0^w C_{n-1}(x')A(x')dx'\,\, .
\end{equation}
These equations are the general version of Eq.~(\ref{eq:ZR1}) and (\ref{eq:tail}) in which $A(x)=0$ when $x>L$. These equations are useful to include non-cylindrical geometries of the ampoule containing the charge. For example to take into account a conic-shaped tip the result is the following:
\begin{equation}
A(x)=A_0\left\{\begin{array}{cl}\left(\frac{x}{X_C}\right)^2 & {\rm if}\,\, x\leq X_C\\ 1 &  {\rm if}\,\, x> X_C\end{array}\right.
\end{equation}
where $A_0$ and $X_C$ are the cross--sectional area of the cylindrical body and the length of the tip, respectively.

\subsection{Case studies}\label{sec:examples}

In Fig.~\ref{fig:distribution_k05_short_ampoule} we show the distribution of impurities for $k=0.5$, which is a value close to the distribution coefficient for potassium in NaI \cite{suerfu2021}, and with an ingot 5 times larger than the molten zone after 1, 5, 10, 15 and 20 passes with and without including the correction for a conic tip. Notice the cuspid at $x=0.8L$ corresponds to the tail of the ingot, where impurities are accumulated for $k<1$. The ultimate distribution is not shown in this figure, as it would practically overlap with the distribution for 20 passes for the ratio $L/w$ considered here. This shows that after $\sim 20$ passes it is not useful to continue the purification process. In the same figure we compare the distribution for a cone-shaped tip with $X_C=0.3w$ (dashed curve). We notice that the effect of taking into account a conic shaped tip is almost negligible, especially when we approach the ultimate distribution. This configuration (with $L/w=5$) was used to check the model and commission the zone refining equipment.

\begin{figure}[b!]
 	\centering
 	\includegraphics[width=0.9\textwidth]{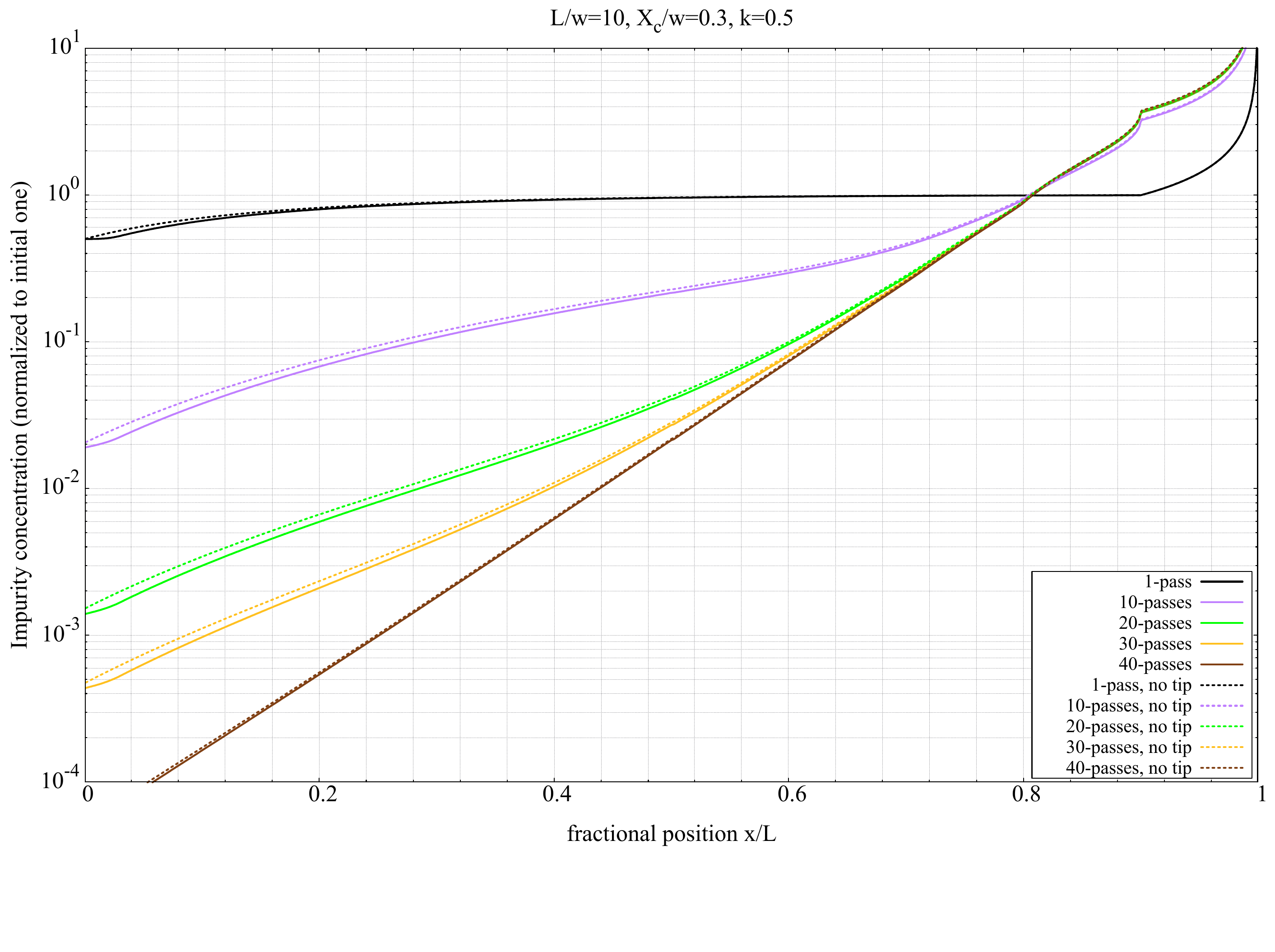}\vspace{-1cm}
 	\caption{Like Fig.~\ref{fig:distribution_k05_short_ampoule} but with $L/w=10$ and for 1, 10, 20, 25 and 40 passes (notice the different scale on y-axes).}
 	\label{fig:distribution_k05_long_ampoule}
 \end{figure}

In Fig.~\ref{fig:distribution_k05_long_ampoule} we do the same exercise but considering a longer ingot $L/w=10$ and for  1, 10, 20, 25 and 40 passes. We notice that with a longer ampoule we can reach a much better purification (notice that the $y$-scale is different from previous figure) increasing the number of passes. We also notice that the effect of the tip is even smaller than in the previous case. For this reason, in the following we neglect the effect of the conical tip and consider for simplicity a cylindrical shaped ingot.

\begin{figure}[t!]
 	\centering
 	\includegraphics[width=0.9\textwidth]{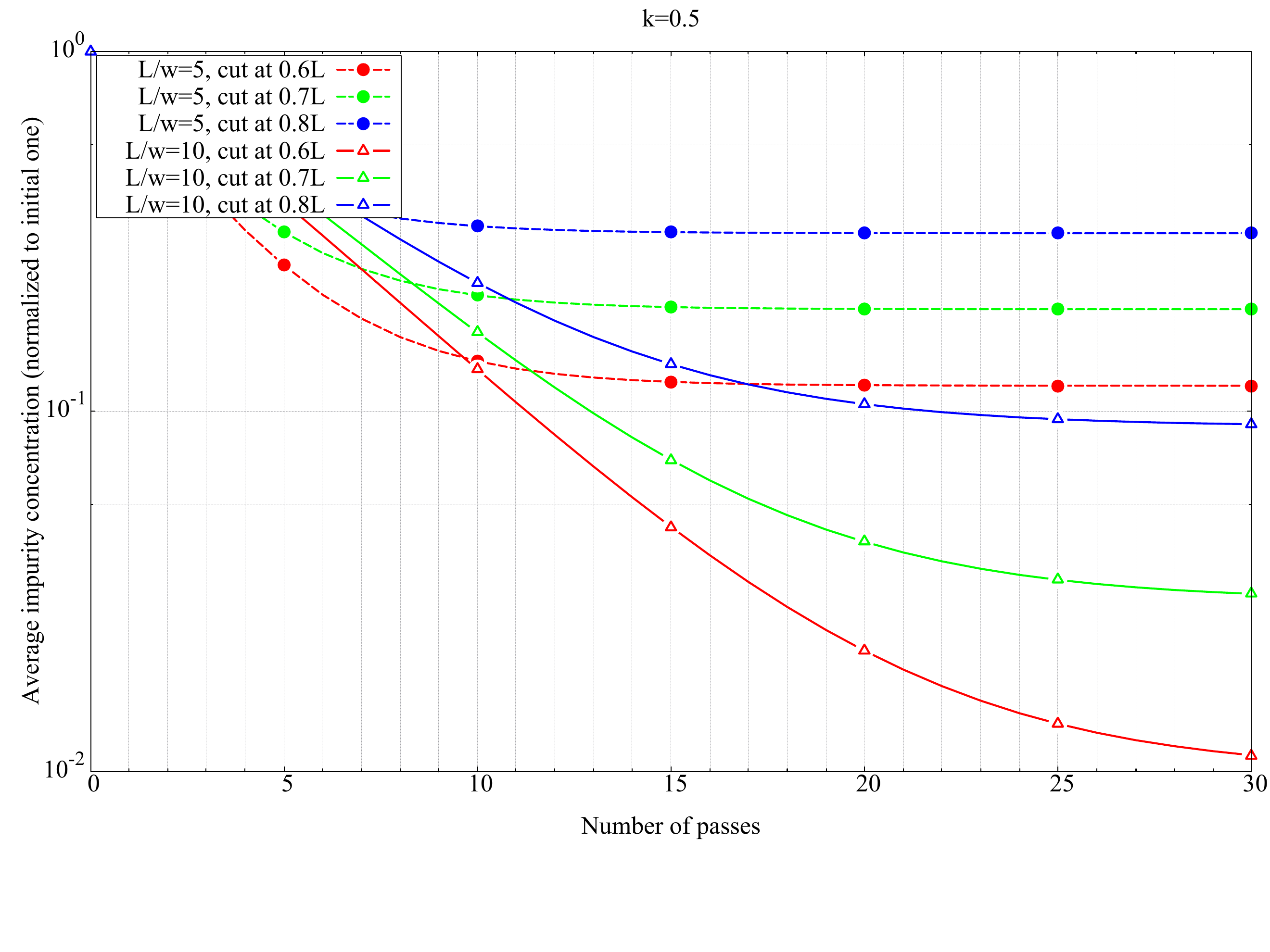}\vspace{-1cm}
 	\caption{Average impurity concentration as function of number of passes for $L/w=5$ and $10$ and for  $X_{\rm cut}/L=0.6$, $0.7$ and $0.8$.}
 	\label{fig:passes}
 \end{figure}

\begin{figure}[t!]
 	\centering
 	\includegraphics[width=0.9\textwidth]{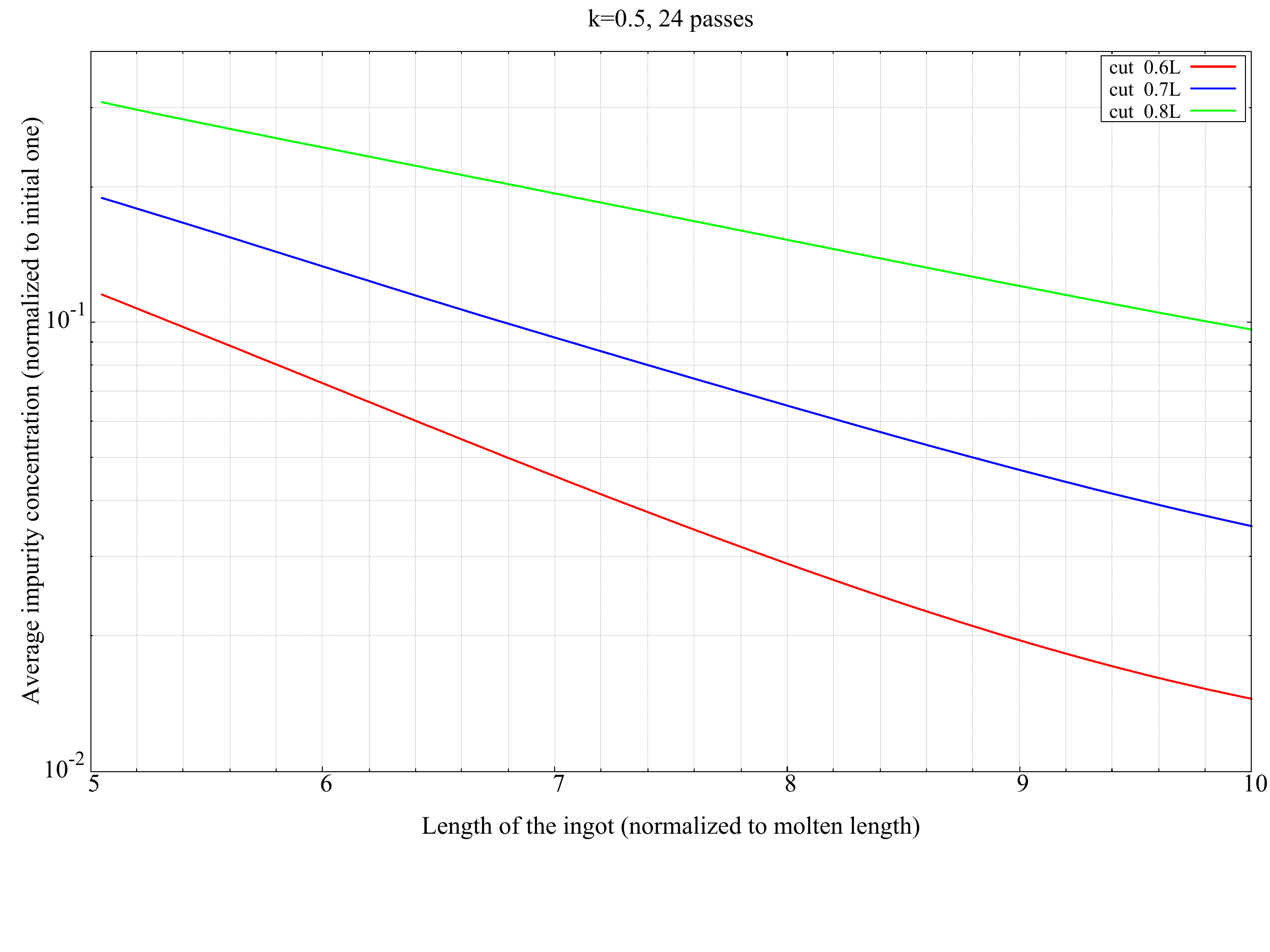}\vspace{-1cm}
 	\caption{Average impurity concentration as function of length of the ingot for $X_{\rm cut}/L=0.6$, $0.7$ and $0.8$ and for 24 passes.}
 	\label{fig:average_vs_L}
 \end{figure}

Usually, after zone refining the tail of the ingot is cut away and the purified part is crunched and than melted again to obtain a pure crystal. For this reason it is useful to calculate the average contamination of the purified part:
\begin{equation}
{\overline C}=\frac{1}{{X_{\rm cut}}}\int_0^{X_{\rm cut}} C_n(x)dx'\,\, ,
\label{eq:cut}\end{equation}
where $n$ is the number of passes. In Fig.~\ref{fig:passes} we show the average impurity concentration as a function of the number of passes both for $L/w=5$ and $L/w=10$ when we cut at $X_{\rm cut}$ of the ingot. As noticed before for $L/w=5$ after $15$--$20$ passes it is not possible to obtain a better purification. A purification of order $10^{-1}$ would be possible at the cost of cutting away as much as 40\% of the ingot. With a longer ingot and 24 passes, the same purification ($10^{-1}$) can be reached by cutting away only 20\% of the ingot, and a deeper purification can be reached by sacrificing a larger fraction of it.
A longer ingot is thus in general more efficient for purification purposes. This is shown in Fig.~\ref{fig:average_vs_L}, where the average impurity concentration is reported as function of the length of the ingot for three values of $X_{\rm cut}$ and for a fixed number of passes (we choose 24 as reference).

\begin{figure}[t!]
 	\centering
 	\includegraphics[width=0.9\textwidth]{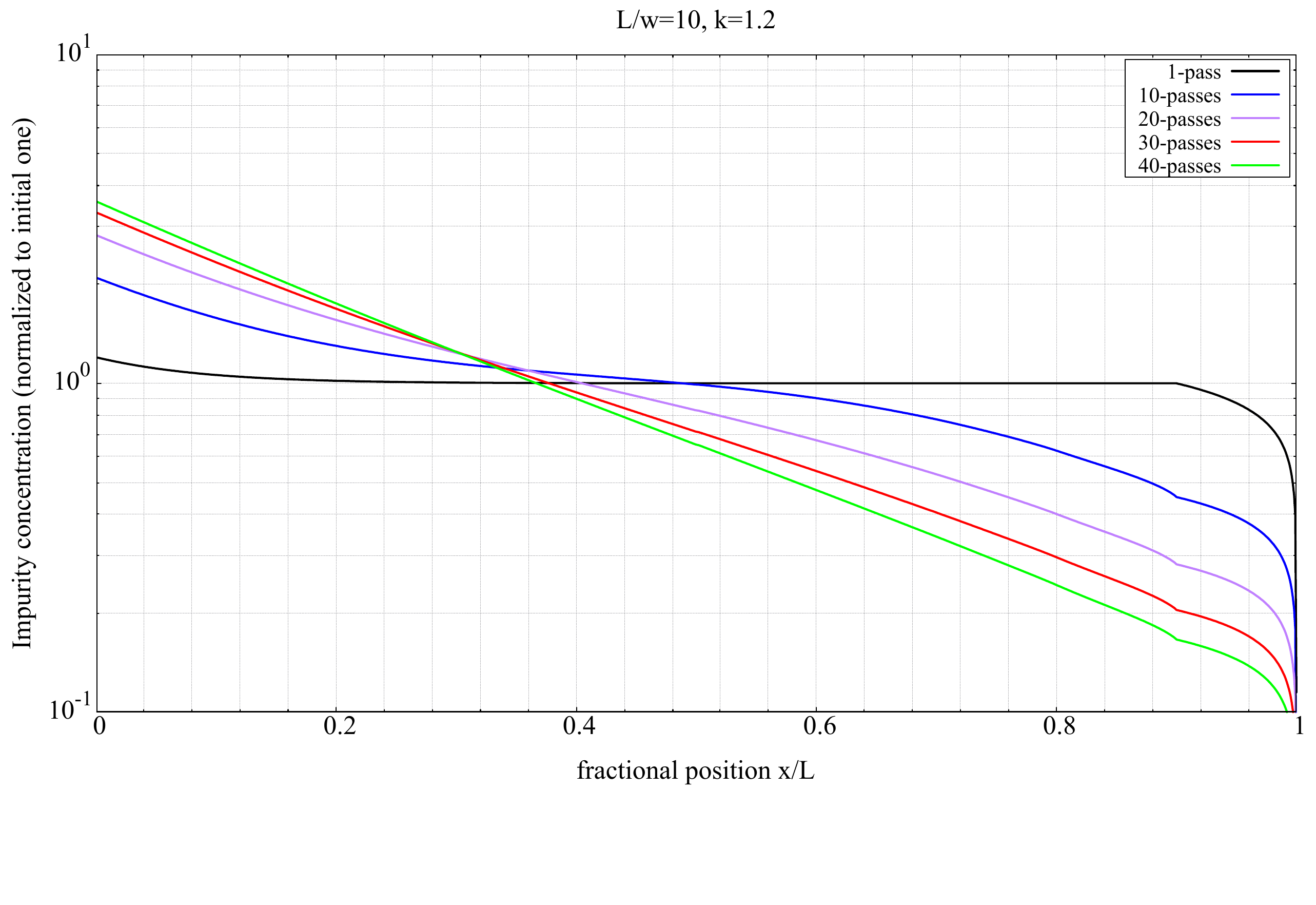}\vspace{-1cm}
 	\caption{Distribution of impurities for $k=1.2$ and $L/w=10$ as function of position $x$ normalized to ingot length $w$ for 1, 10, 20, 30 and 40 passes.}
 	\label{fig:distribution_k12}
 \end{figure}

\begin{figure}[t!]
 	\centering
 	\includegraphics[width=0.9\textwidth]{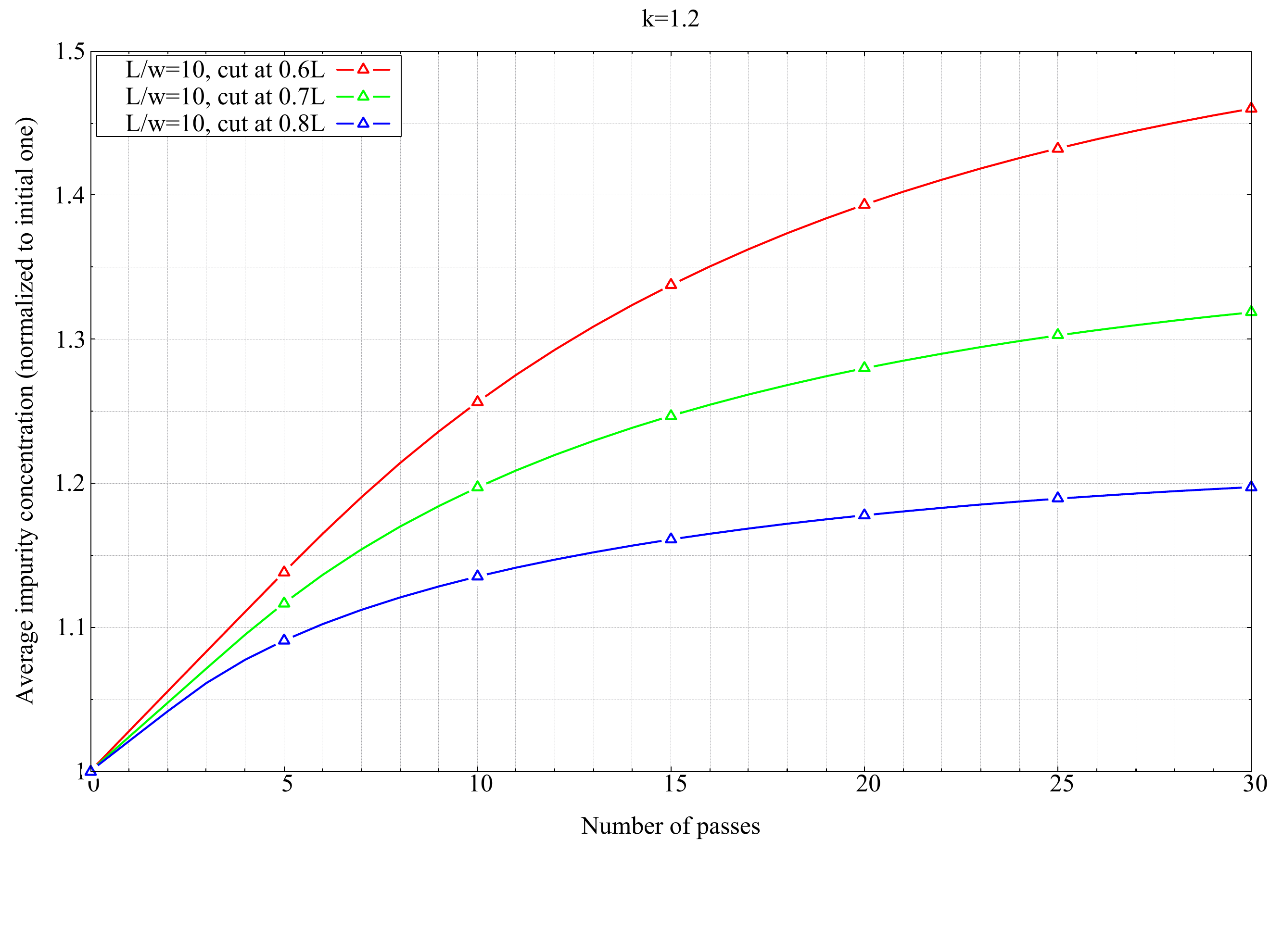}\vspace{-1cm}
 	\caption{Average impurity concentration as function of the number of passes for $L/w=10$ and for  $X_{\rm cut}/L=0.6$, $0.7$ and $0.8$}
 	\label{fig:k12_passes}
 \end{figure}

In Fig.~\ref{fig:distribution_k12} we show the distribution of impurity for $k>1$ (namely $k=1.2$). In this case we have the opposite effect: since impurities prefer to move to the frozen zone they tend to accumulate in the tip of the ingot. This can represent a problem  because when one tries to confine impurities with $k<1$ in the tail of the ingot at the same time contaminants with $k>1$ move towards the tip. However, from Fig.~\ref{fig:k12_passes} we can see that the increase in the impurity concentration is modest when $k \sim 1$.

\section{Experimental procedure}\label{SecIV}

\begin{figure}[t!]
    \centering
    \includegraphics[width=0.8\textwidth]{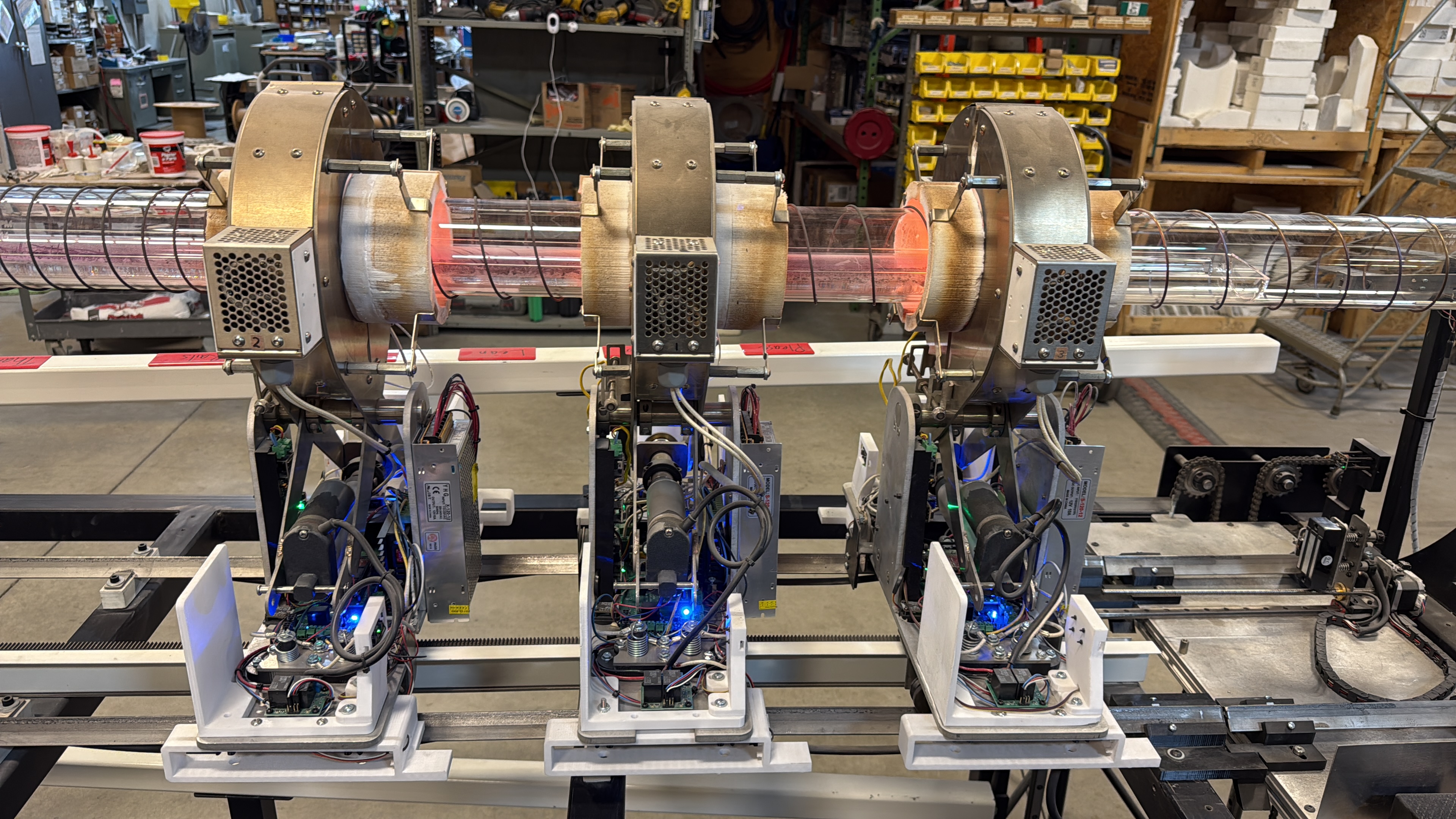}
    \caption{The refurbished Wavefront zone refiner in operation. The loaded crucible is positioned inside a quartz "boat" (visible on the right), both of which are housed inside a quartz tube with a heating coil.}
    \label{fig:zonerefiner}
\end{figure}

A primary challenge in handling NaI is its high hygroscopicity. Consequently, the powder must be processed in an inert atmosphere, typically inside a glovebox filled with nitrogen and maintained at ppm-level residual moisture. To minimize external contamination, synthetic quartz crucibles (or ampoules) are used for zone refining.
The ampoule preparation is conducted at Radiation Monitoring Devices (RMD) in Watertown, MA, where the final crystal growth also takes place. Since molten NaI tends to stick to quartz, which can lead to tube cracking upon cooling, we utilized two methods to prevent this from happening: 1) carbon coating of the ampoule inner surface and 2) SiCl$_4$ vapor treatment prior to refining \cite{eckstein1968}. In method 1), ampoules are procured directly from an external company with the coating applied. In method 2), we use the following procedure: the dried powder charge is placed in a quartz ampoule inside a furnace and exposed to SiCl$_4$ vapor at 400~$^\circ$C for 30 minutes, then the temperature is reduced and the ampoule is sealed. Residual SiCl$_4$ vapor may either be removed prior to sealing or retained inside the sealed ampoule. 
The final step before zone refining involves melting the powder horizontally to create a solidified charge of uniform thickness. This procedure ensures the solid ingot remains loose inside the ampoule. 
We notice that alternative methods to SiCl$_4$ exist \cite{Gross1970}.
Following treatment, the crucible is transported to the Mellen Company in Concord, NH, for processing in a three-heating coil horizontal Wavefront zone refiner. 
The zone refiner equipment features a two-level design that enables the three heating coils to operate continuously \cite{SuerfuPhD}. Zone refining takes place at the upper level where the ampoule is housed in pre--heated quartz tube. At the end of the run each heating coil is brought to the lower level by a lift and then rapidly moved back to the initial position allowing for uninterrupted multiple pass operation.  The heating coils speed can be tuned  to optimise the purification process. 
 The zone refiner was initially used in 2015 \cite{suerfu2021} in the framework of the SABRE project. In 2023 it underwent extensive refurbishment in order to facilitate the mass production required for SABRE North. Figure~\ref{fig:zonerefiner} shows the refurbished  zone refiner in operation.
The upgrades enhanced nearly every major component of the Wavefront zone refiner—including the structural frame, motion mechanisms, and control systems—to improve reliability and precision.

A crucial upgrade was the replacement of the system's central processing units (CPUs). The new CPUs provide expanded input capacity, enabling more sophisticated control over translation and cart-to-cart interactions. Communication protocols were upgraded from TCP/IP (call-and-response) to User Datagram Protocol (UDP), significantly reducing latency and increasing data streaming speed. While the original frame utilized a dual-chain setup for translation, the new configuration employs independent stepper motors for each furnace cart, integrated with a rack-and-pinion assembly. This allows for independent travel speeds: processing 
can be as slow as 0.64~cm per hour, while rapid travel between cycles can reach up to 17.8~cm per minute. Furthermore, the original scissor lifts were replaced with stable elevator lifts featuring electromagnetic clamps. 

The sensor suite was also upgrades. The original infrared (IR) sensors, which were sensitive to ambient lighting and shadows, were replaced with laser-type sensors. These provide consistent measurements regardless of ambient light and offer a wider focal range. Additional laser sensors were installed on the elevator lifts to confirm cart loading and on the front and rear of each furnace cart to monitor inter-cart distances. The precise spacing between the carts allows control of the temperature gradient, resulting in a steeper profile as the cart separation increases. These modifications have collectively improved the stability and control of the entire purification system. 

To validate our mathematical model and determine the distribution coefficients for various contaminants, we conducted a series of experimental runs, categorized into commissioning and production phases. The commissioning runs tested different crucible treatments and configurations. We used a 61~cm crucible, corresponding to a length-to-zone-width ratio ($L/w$) of 5, with a 0.9~kg charge. We tested three options: carbon-coated crucibles and SiCl$_4$ treatments in sealed crucibles both with and without residual gas.

For the production runs, we utilize a 122~cm crucible ($L/w = 10$) with a 2~kg charge, employing the SiCl$_4$ treatment with residual gas. This was preferred, time and cost wise, to the coating, given the positive outcome of the commissioning runs.
The $L/w = 10$ ratio was selected based on our mathematical model to reduce the concentration of $^{40}$K—the primary background source 
in the ROI—by a factor of 10 while retaining 80\% of the charge. Under these conditions, five zone-refining runs are required to produce enough  
material for a 5~kg crystal. Investigations into increasing the charge capacity are currently ongoing.

\section{Results and data analysis}\label{SecV}

\subsection{Commissioning runs}

 We analysed data from one among the commissioning runs of the zone refiner, defined as RUN1. A 61~cm length, 3.6~cm inner diameter, and 2~mm thick carbon coated synthetic quartz ampoule was used to refine 0.9~kg of Astro Grade powder with 26 passes at a speed of 3.8~cm/h. After the process is complete, the ingot is cut at 2.5, 15, 30.5, 45.8, and 58.4 cm distance from the tip. Then these slices are crunched and samples are taken for ICP-MS measurements. The results for six contaminants are shown in Table~\ref{tab:measures}. Besides $^{39}$K, $^{208}$Pb, and $^{85}$Rb, which are relevant for estimating the background in SABRE North, we consider other four contaminants for the purpose of testing our mathematical model. In the table we compare the concentration of the contaminants at different positions in the ingot with the initial concentration in the powder (first row). For some measurements the contaminant concentration is below the sensitivity of the instrument which is therefore reported as upper limit. Hereafter, we refer to this data as RUN1 data.

\begin{table*}[t!]
\centering
\scalebox{1}{
\begin{tabular}{c|ccccccc}
\hline
            & $^{138}$Ba  & $^{44}$Ca   & $^{39}$K    & $^{24}$Mg   & $^{208}$Pb  & $^{88}$Sr  & $^{85}$Rb\\
\hline
Powder      & 0.18$\pm$0.4 & $<$100     & 6.5$\pm$1.6 & 2.1$\pm$0.5 & 1.6$\pm$0.3 & unknown  & $<0.4$  \\
\hline
Sample position & & & & & & & \\ 
2.5$\pm$1.3 & $<$0.3         & 183$\pm$37 & $<$4        & 2.4$\pm$0.5 & 3.6$\pm$0.7 & 367$\pm$74  & $<0.4$ \\
15.2$\pm$1.3 & $<$0.3       & 147$\pm$30 & $<$4        & 1.9$\pm$0.6 & 2.9$\pm$0.6 & 287$\pm$58  & $<0.4$ \\
30.5$\pm$1.3& $<$0.3       & 41$\pm$10  & $<$4        & 2.0$\pm$0.4 & 1.8$\pm$0.15& 86$\pm$17  & $<0.4$ \\ 
45.8$\pm$1.3& $<$0.3       & 21$\pm$5   & 6.4$\pm$1.2 & 8.3$\pm$2.2 & 1.2$\pm$0.2 & 41$\pm$8    & $<0.4$ \\
58.4$\pm$1.3& 4.0$\pm$0.8  & $<$16      & 270$\pm$17  & 10.4$\pm$2.4& 0.9$\pm$0.2 & 10$\pm$2    & $<0.4$ \\
\hline
\end{tabular}}
\caption{Concentration from ICP-MS for six different contaminants (in ppb) for the Astro Grade powder before and after zone refining in RUN1. Five samples from the zone refined ingot are taken at different positions (in cm) from the tip to the tail. The errors are only statistical (r.m.s. on three different measurements).  A 20\% systematic error is added in quadrature to all the measurements.\label{tab:measures}}
\end{table*}

\subsubsection{Statistical analysis}
In order to estimate the distribution coefficient for the various contaminants we perform a statistical analysis. We have to face three main problems: a) some samples present only upper limits; b) samples are taken randomly from the crunched slices, so we have an uncertainty on the position of the measurement; c) the systematic error on the measurement is unknown. For the last point we assume conservatively a systematic error of 20\% on measurements above the sensitivity threshold (Limit of Quantification, LoQ) to add in quadrature to the statistical one. In absence of further information we assume systematic errors as uncorrelated. 

We build a likelihood function defined as follows:
\begin{equation}
L(k,C_0)=\prod_{z=1}^5{\cal P}_z(k,C_0)\,\, ,
\end{equation}
where $k$ and $C_0$ are the (unknown) distribution coefficient and initial powder uniform impurity concentration, and the probability functions ${\cal P}_z(k,C_0))$ are defined as follows:
\begin{itemize}
\item For a measurement above the detection limit
\begin{equation}
{\cal P}_z(k,C_0)\propto\int_{x_z-\frac{\delta}{2}}^{x_z+\frac{\delta}{2}}
\exp\left[-\frac{\left(C_z^{\rm exp}-C_0 f(k,x)\right)^2}{2\sigma_z^2}\right]\, dz\,\, ,
\end{equation}
where $x_z$ is the central position of the sample $z$, $\delta$ the dimension of the slice,  $C_z^{\rm exp}$ is the measured value of concentration, $f(k,x)$ is the theoretical distribution after $n=26$ passes with initial concentration $C_0$, $\sigma_z$ the error of the measurement. In this way the uncertainty on the position of the grain in the bin is taken into account by integrating on the bin size.

\item For a measurement in which only an upper limit is given, ${\cal P}_z$ is calculated as the fraction of the interval $[x_z-\frac{\delta}{2},x_z+\frac{\delta}{2}]$ in which $C_0 f(k,x)<C_z^{\rm LoQ}$. In fact, if the concentration exceeded the LoQ, we would have a measurement and not an upper limit. In this way,${\cal P}_z=1$ (${\cal P}_z=0$) if $C_0 f(k,x)>C_z^{\rm LoQ}$ ($\leq C_z^{\rm LoQ}$) in the interval $[x_z-\frac{\delta}{2},x_z+\frac{\delta}{2}]$. For example for $k<1$ and thus $f(k,x)$ increasing function, if there is a point  in which $C_0 f(k,X_C)=C_z^{\rm LoQ}$ we have ${\cal P}_z=(X_C-x_z+\delta/2)/\delta$.

\end{itemize}

To avoid that the fitted value of $C_0$ deviates too much from the value measured in the powder, we multiply the likelihood function by a prior
\begin{equation}
L_P(k,C_0)=L(k,C_0)\times\exp\left[-\frac{\left(C_0^{\rm powder}-C_0\right)^2}{2\sigma_{\rm powder}^2}\right] \,\, ,
\end{equation}
where we assume $\sigma_{\rm powder}=0.2 C_0^{\rm powder}$. Finally, we define the likelihood ratio functions
\begin{eqnarray}
LR(k,C_0)&=&-2\,\log\left[\frac{L(k,C_0)}{L({\hat k},{\hat C}_0)}\right]\, ,\nonumber\\
LR_P(k,C_0)&=&-2\,\log\left[\frac{L_P(k,C_0)}{L_P({\hat k},{\hat C}_0)}\right]\,\, ,
\end{eqnarray}
where ${\hat k}$ and ${\hat C}_0$ are the values that maximize the likelihood functions (best fit values). According to the Wilk’s theorem this function behaves approximately like a $\chi^2$.

\subsubsection{Results from commissioning run}
 
\begin{figure}[t!]
 	\centering
 	\includegraphics[width=1.\textwidth]{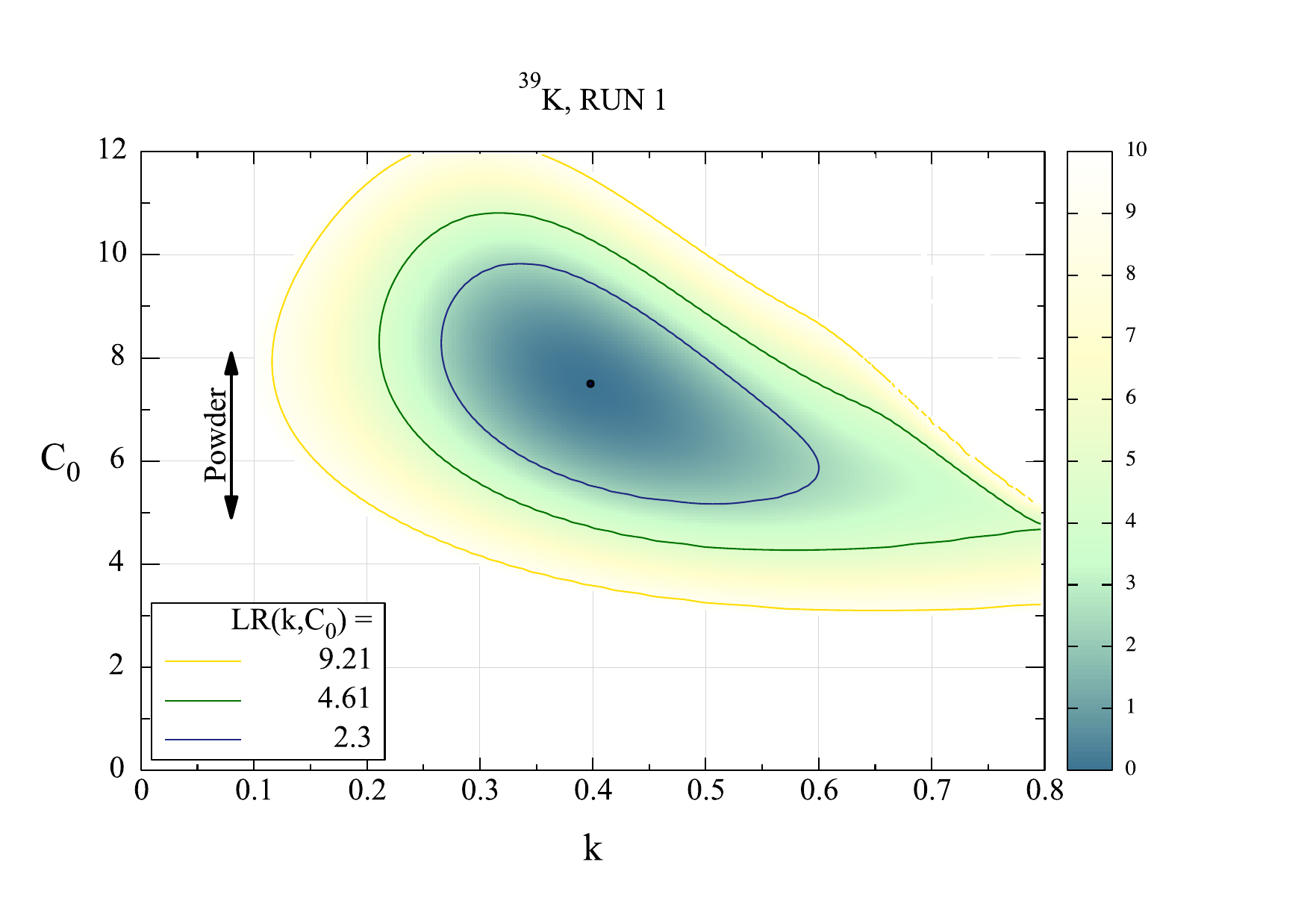}\vspace{-0.5cm}
 	\caption{Allowed zones in the parameter space $(C_0,k)$ for the $^{39}$K contaminant. The blue, green and yellow lines correspond to $LR(K,C_0)=2.30$, $4.61$ and $9.21$ (i.e., 1, 2 and 3$\sigma$ C.L.). The arrow is the $1\sigma$ measurement for $C_0$ in the powder, while the blue dot is the best fit value. The prior on $C_0$ from the powder measurement has been used in the analysis.}
 	\label{fig:K39_Allowed}
 \end{figure}

A fit to the data in Table~\ref{tab:measures} is performed, seperately for each contaminant, using the likelihood function described in the previous section. In Fig.~\ref{fig:K39_Allowed} the allowed zone in the parameter space $(C_0,k)$ for the $^{39}$K contaminant is shown. The blue, green and yellow lines correspond to $LR(K,C_0)=2.30$, $4.61$ and $9.21$ (corresponding to 1, 2 and 3$\sigma$). The best fit is found for $C_0=7.5{\rm\, ppm}$ (which is compatible with the value measured in the powder, $C_0=6.5\pm 1.6{\rm\, ppm}$) and $k=0.4$. The $1\sigma$ interval for $k$ can be obtained marginalizing the likelihood ratio with respect to $C_0$. 

 \begin{figure}[b!]
 	\centering
 	\includegraphics[width=0.9\textwidth]{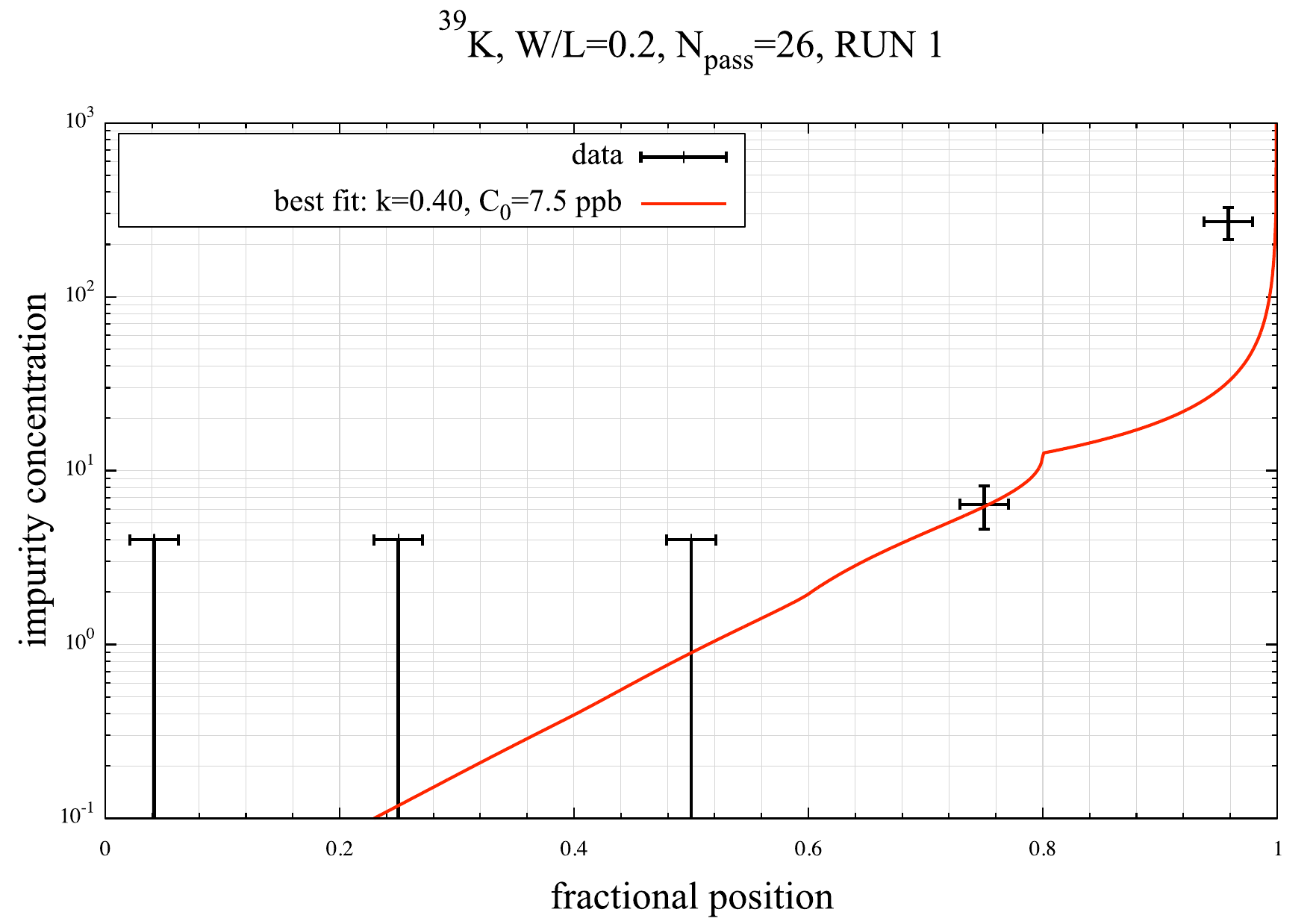}\vspace{0cm}
 	\caption{Best fit distribution for the $^{39}$K contaminant. The impurity concentration is given in ppb.}
 	\label{fig:K39_Best}
 \end{figure}

In Fig.~\ref{fig:K39_Best} we show the $^{39}$K distribution for the best fit values of $k$ and $C_0$. Since the first three bins are upper limits, the prior on powder is necessary in the analysis. Without the prior, very steep distributions can fit the last two bins as well, allowing unnaturally low values of $k$, which correspond to values of $C_0$ much larger than those actually measured in the powder.  We have also verified that excluding the last bin (where the concentration can vary significantly with the specific sampling location) the results are almost unchanged.

\begin{figure}[t!]
 	\centering
 	\includegraphics[width=1.\textwidth]{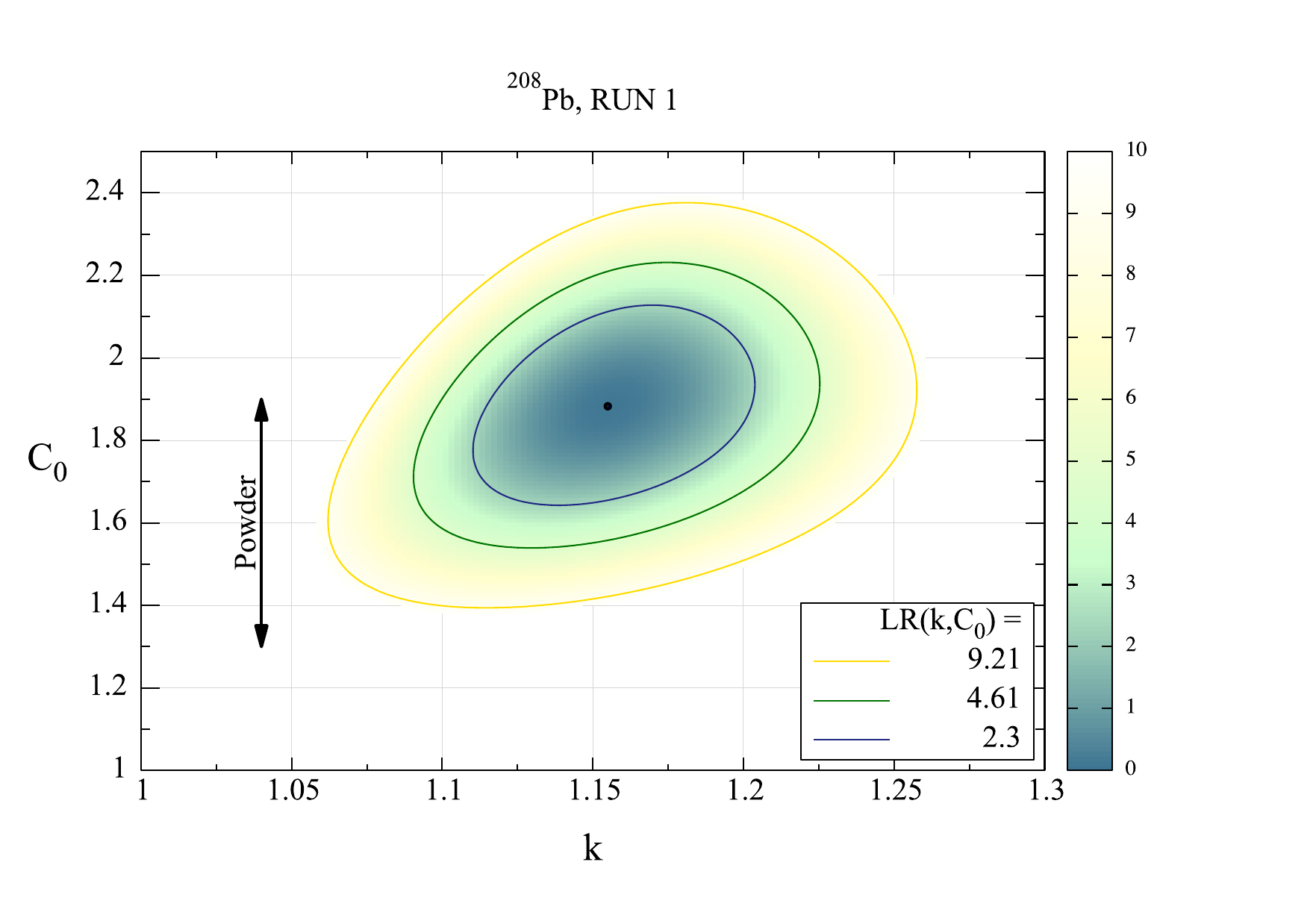}\vspace{-0.5cm} 	\caption{Same as Fig.~\ref{fig:K39_Allowed} but for the $^{208}$Pb contaminant. Note that the scale and the ranges are different from Fig.~\ref{fig:K39_Allowed}.}
 	\label{fig:Pb208_Allowed}
 \end{figure}

\begin{figure}[b!]
 	\centering
 	\includegraphics[width=0.9\textwidth]{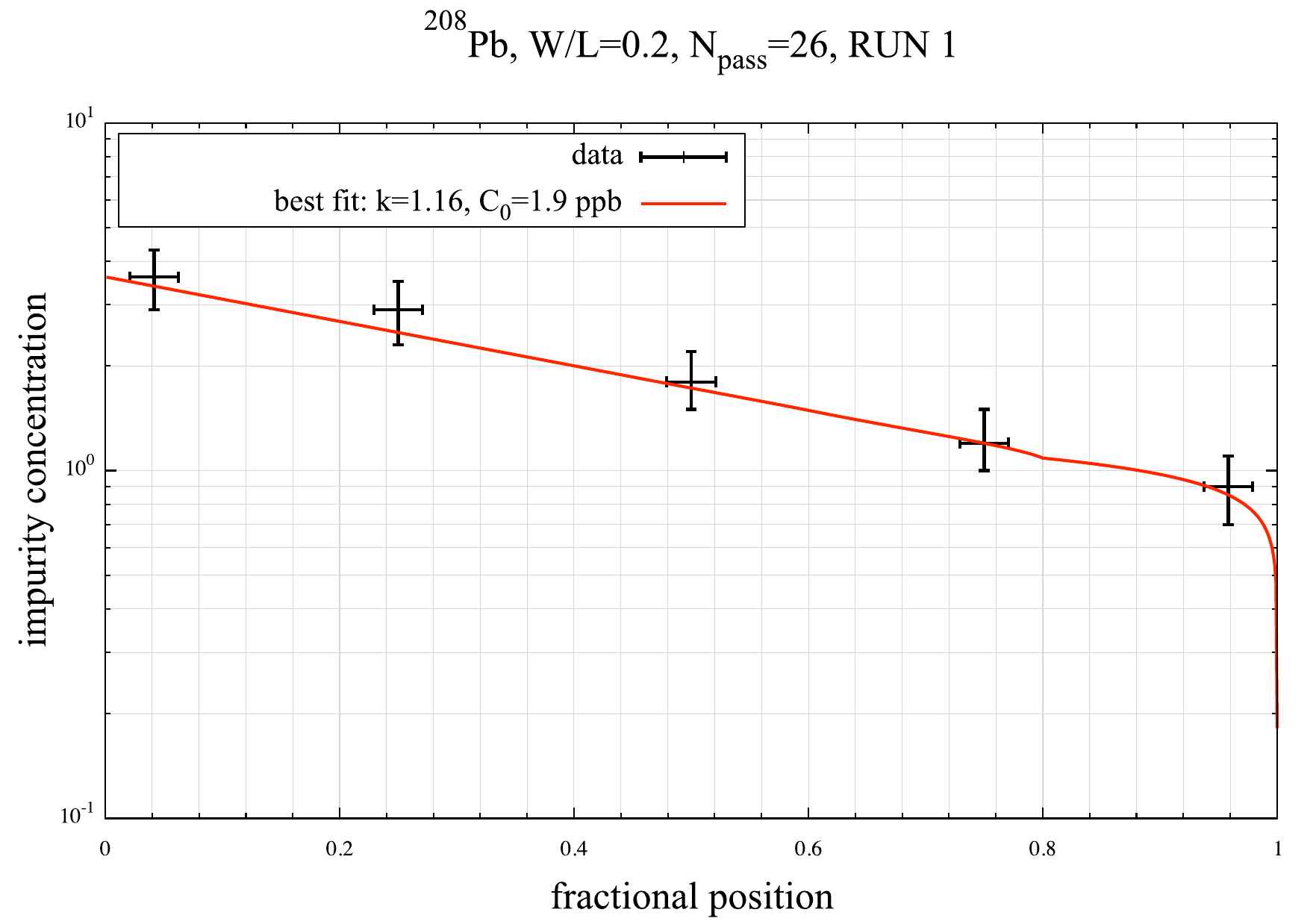}\vspace{0cm}
 	\caption{Best fit distribution for the $^{208}$Pb contaminant. The impurity concentration is given in ppb.}
 	\label{fig:Pb208_Best}
 \end{figure}

In Fig.~\ref{fig:Pb208_Allowed} we show the allowed zone for the $^{208}$Pb contaminant. In this case we have $k>1$ and thus the distribution is decreasing. This can be seen in Fig.~\ref{fig:Pb208_Best} where the distribution for the best fit is shown. Since we have a measurement for all five bins we don't need to constrain the value of $C_0$ with the prior on the powder contamination. In fact, we verified that, by adding or excluding the prior, compatible results are obtained. We see that in this case $^{208}$Pb tends to remain preferably in the tip of the ingot, thus contaminating the zone purified from $^{39}$K. However, we see that the contamination is modest (a factor 1.13 on average with respect to the powder initial contamination retaining 80\% of the ingot).

We have repeated the fit for the six contaminants listed in  Table~\ref{tab:measures}. For $^{85}$Rb the measures are always below the Limit of Quantification (LoQ) and thus are not useful to extract values of $k$ and $C_0$. We show the results in Table~\ref{tab:results}. For the $^{138}$Ba and $^{39}$K the addition of the prior on $C_0$ is necessary since the first bins are upper limits. For $^{44}$Ca we have only an upper limit on $C_0<100$ ppm in the powder. In this case the prior function is simply $\Theta(100-C_0)$, where $\Theta$ is the Heaviside function. In any case we have verified that the results are  independent form the prior, and the fitted value of $C_0$ is 68.7, compatible with the upper limit on powder. For $^{88}$Sr we have excluded the value coming from the powder because it is not known. 

\begin{table*}[t!]
\centering
\scalebox{0.9}{\begin{tabular}{c|c|cccccc}
\hline
& & $^{138}$Ba  & $^{44}$Ca   & $^{39}$K    & $^{24}$Mg   & $^{208}$Pb  & $^{88}$Sr  \\
\hline
Powder  & $C_0$ (ppb)     & 0.18$\pm$0.4 & $<$100     & 6.5$\pm$1.6 & 2.1$\pm$0.5 & 1.6$\pm$0.3 & unknown \\
\hline\\[-1.8ex]
\multirow{2}{5em}{with prior} 
& $k$ & 0.34$\pm$0.15 & 1.36$\pm$0.05 & 0.40$^{+0.09}_{-0.12}$  &0.89$\pm$0.07 &1.16$\pm$0.03 & --\\
& $C_0$ (ppb) & 0.21 & 68.7 & 7.5 & 2.6 & 1.9 & -- \\
\hline 
\multirow{2}{5em}{w/o prior} 
& $k$ & -- & 1.36$\pm$0.05 & -- & 0.82$\pm$0.07 & 1.16$\pm$0.03 & 1.42$\pm$0.03\\
& $C_0$ (ppb) & -- & 68.7 & -- & 3.3 & 2.0 & 135\\
\hline
\end{tabular}}
\caption{Fitted values of the distribution coefficient $k$ and the initial contaminant concentration $C_0$ in a selected commissioning run with $L/w=5$. \label{tab:results}}
\end{table*}

\subsection{Production runs}\label{sec:PR}

\begin{table}[h!]
\centering
\begin{tabular}{c|ccccccc}
\hline
             & $^{138}$Ba  & $^{44}$Ca   & $^{39}$K    & $^{24}$Mg   & $^{208}$Pb  & $^{88}$Sr & $^{85}$Rb\\
\hline 
Sample position & & & & & & & \\
15.2$\pm$1.3  & 0.24   & 62.8 & $<$4.7 & 5.0  & 2.0 & 1.50 & $<$0.5\\
45.8$\pm$1.3 & $<$0.1 & 29.0 & $<$4.7 & $<$2 & 2.0 & 0.31 & $<$0.5\\
76.2$\pm$1.3 & $<$0.1 & 18.0 & $<$4.7 & $<$2 & 2.0 & 0.40 & $<$0.5\\ 
106.7$\pm$1.3 & 5.30   & $<$7   & 55     & 4.0  & 2.0 & $<$0.06 & $<$0.5\\
\hline
\end{tabular}
\caption{Concentration from ICP-MS for six different contaminants (in ppb) for the Astro Grade powder before and after zone refining in RUN2. Four samples from the zone refined ingot are taken at different positions (in cm) from the tip to the tail. No statistical errors are provided.  A 20\% systematic error is estimated for each measure. \label{tab:measure2}}
\end{table}

After the commissioning of the zone refiner we started the production of zone refined powder for SABRE North crystals growth. We report our study of one selected production run, defined as RUN2. In this case a 122-cm length, 3.6~cm inner diameter, and 2~mm thick synthetic quartz ampoule was used to refine 2~kg of Astro Grade powder with 24 passes at a speed of 5~cm/h. After refining, the full ingot has been divided into four sections, each 30.5 cm long. At the middle of each section a sample for ICP-MS has been taken. The powder batch used in this run was already assayed for previous purposes. The results obtained from the RUN2 are shown in Table~\ref{tab:measure2} for the same contaminants reported in Table~\ref{tab:measures}. For $^{138}$Ba we have a measure in the first and last bin while the central bins are only upper limits. These data do not allow to fix a single value for $k$ since distributions with both positive or negative slopes are allowed. For this reason we do not use this data to fit the value of $k$ for $^{138}$Ba. We also notice that the value of $C_0$ for $^{88}$Sr is completely different (by two order of magnitudes) from the values measured in RUN1. This may be due to a contamination of $^{88}$Sr in the measurements of RUN1. Indeed, samples in RUN1 where handled in a glove box not only used for SABRE North activities. On the contrary for RUN2 a new glove box has been used, designed to handle longer ampoules and dedicated only to SABRE North. Despite this difference, the value of $k$ is compatible with those measured in RUN1.

\begin{table}[h!]
\centering
\begin{tabular}{c|ccccccc}
\hline
            & $^{44}$Ca              & $^{39}$K              & $^{24}$Mg               & $^{208}$Pb  & $^{88}$Sr   \\
\hline
\\[-1.8ex] 
$k$        & $1.23^{+0.03}_{-0.02}$ & $0.65^{+0.06}_{-0.11}$ & $1.0^{+0.04}_{-0.02}$  &1.0$\pm$0.03 & $1.32^{+0.06}_{-0.02}$ \\[-1.8ex] \\
$C_0$ (ppb) &  29.6                  & 6.3                    & 2.0                    & 1.9         & 0.46        \\
\hline
\end{tabular}
\caption{Fitted values of the distribution coefficient $k$ and the initial contaminant concentration $C_0$ from RUN2 data. The prior on powder is included only for $^{39}$K. \label{tab:results2}}
\end{table}

\subsection{Normal Freezing}

After the zone refining procedure, the purified part of the ingot will be crumbled and recast in a single crystal through the \emph{normal freezing technique} or \emph{Bridgman method}. This is done by melting the NaI in a vertical cylindrical crucible which is extracted very slowly ($\sim$ 1 cm/day) from the heating coil. The procedure is similar to that of the zone refining close to the final part of the ingot and indeed the distribution of contaminants follow the same distribution as in Eq.~(\ref{eq:tail}) with $w=L$ where $L$ is the length of the ingot. 
\par\indent
We have produced a crystal from the same powder used in RUN1 and RUN2. The grown ingot is a 25~cm high cylinder with a cone-shaped tip with $X_c=$7~cm. After growth this crystal was broken in chunks to create a second one. From broken parts we have selected three 0.5 cm thick slices at the bottom of the cylinder ($x_1=7$~cm from the tip), in the middle of the ingot ($x_2=18.5$~cm) and in the tail ($x_3=30$~cm). As in the case of zone refining, the slices are crunched and samples are taken for ICP-MS measurements. The results are shown in Table~\ref{tab:measures_NF}

\begin{table*}[t!]
\centering
\scalebox{1.}{\begin{tabular}{c|ccccccc}
\hline
 Position             &  $^{138}$Ba   & $^{44}$Ca   & $^{39}$K     & $^{24}$Mg     & $^{208}$Pb    & $^{88}$Sr & $^{85}$Rb \\
\hline
7 $\pm$0.5 cm & 0.19$\pm$0.01 & $<$100      & 4.56$\pm$0.32  & 1.60$\pm$0.02 & 1.84$\pm$0.02 & 96.9$\pm$2.2 & $<$0.4\\
18$\pm$0.5 cm & 0.26$\pm$0.02 & $<$100      & 5.79$\pm$1.14  & 1.79$\pm$0.18 & 1.91$\pm$0.05 & 59.0$\pm$1.1 & $<$0.4\\
30$\pm$0.5 cm & 0.46$\pm$0.03 & $<$100      & 9.84$\pm$0.65  & 1.86$\pm$0.85 & 1.94$\pm$0.07 & 22.7$\pm$0.5  & $<$0.4\\ 
\hline
\end{tabular}}
\caption{Concentration from ICP-MS for six different contaminants (in ppb) for the Astro Grade powder after normal freezing crystal growth. Three samples from the zone refined ingot are taken at different positions from the tip. The errors are only statistical (r.m.s. on three different measurements). A 20\% systematic error is added in quadrature to all measurements. \label{tab:measures_NF}}
\end{table*}

 \begin{figure}[b!]
 	\centering
 	\includegraphics[width=0.9\textwidth]{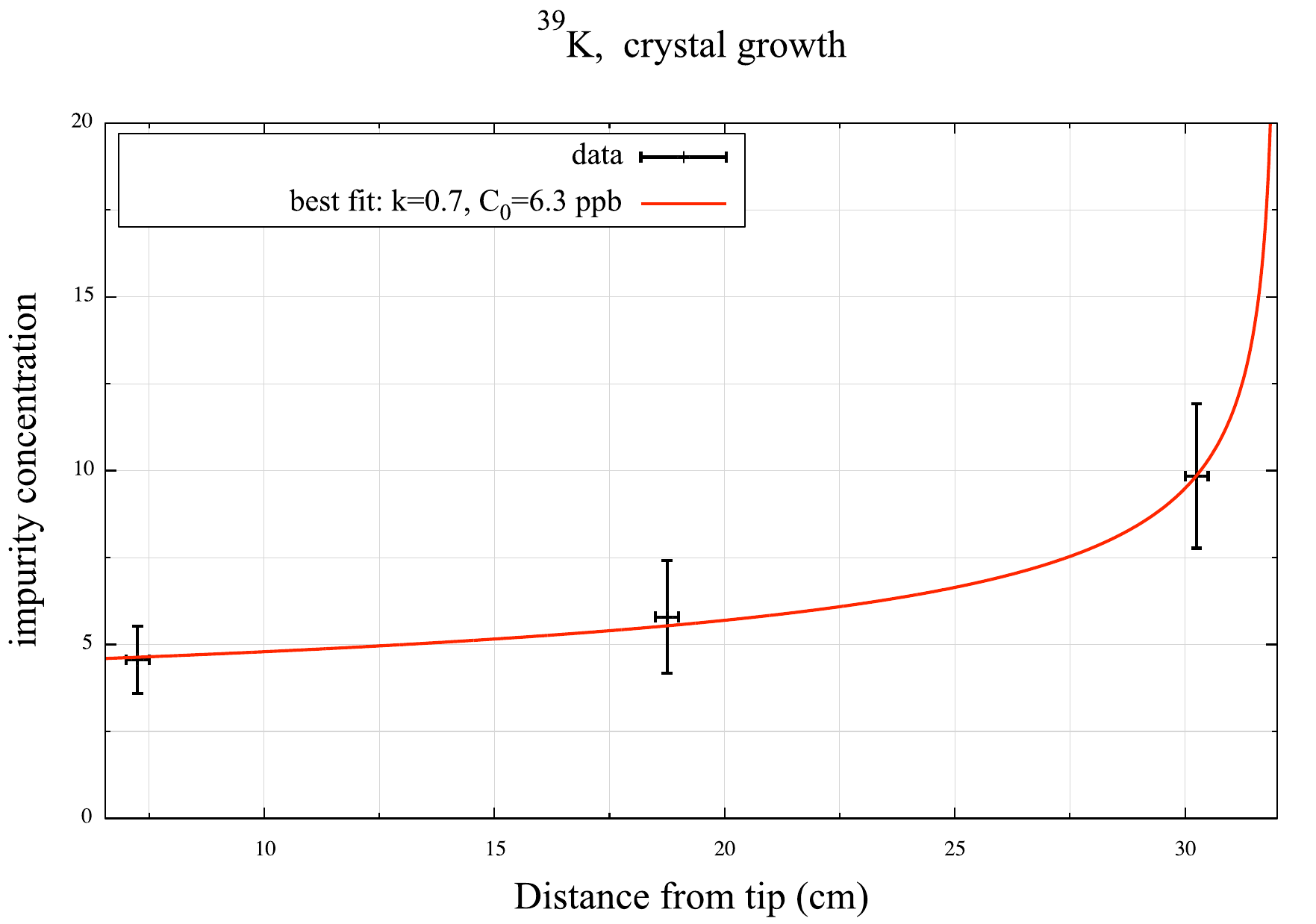}\vspace{0cm}
 	\caption{Best fit distribution for the $^{39}$K contaminant from normal freezing data. The impurity concentration is given in ppb.}
 	\label{fig:K39_Best_NF}
 \end{figure}

Taking into account the geometry of the crucible, the expected distribution of contaminants is
\begin{equation}
C(x)=\frac{kC_0}{\left(1-\frac{2X_c}{3L}\right)^{k-1}}\left(1-\frac{x}{L}\right)^{k-1}\,\,\,\, {\rm for}\,\,\,\, X_x<x\leq L\,\, .\label{eq:tail1}
\end{equation}
where the term in the denominator is due to the cone-shaped tip of the crystal. Fitting this distribution with the data in Table~\ref{tab:measures_NF} we obtain the results in Table~\ref{tab:results_NF}. We notice that the fitted value for $C_0$ for $^{88}$Sr is again different from those obtained in RUN1 and RUN2. For $^{44}$Ca we have only upper limits in the data and for this reason $k$ is totally unconstrained.

\begin{table}[h!]
\centering
\begin{tabular}{c|cccccc}
\hline
& $^{138}$Ba  & $^{39}$K    & $^{24}$Mg               & $^{208}$Pb   & $^{88}$Sr  \\
\hline\\[-1.8ex]
$k$ & 0.68$\pm$0.11 &0.72$\pm$0.11 & $0.93^{+0.23}_{-0.17}$ &0.98$\pm$0.1 & 1.53$\pm$0.11 \\
$C_0$ (ppb) & 0.28 &  6.3           & 1.7                     & 1.9          & 61.5        \\
\hline
\end{tabular}
\caption{Fitted values of the distribution coefficient $k$ and the initial contaminant concentration $C_0$ from normal freezing crystal growth data.\label{tab:results_NF}}
\end{table}

In Fig.~\ref{fig:K39_Best_NF} we show the distribution for the best fit values obtained for $^{39}$K from normal freezing data. We notice that there are not substantial differences between the values of $k_{eff}$ found with normal freezing and those found with zone refining. 
Since the crystallization front in normal freezing advances much more slowly than the heating coil during zone refining, this ensures that in zone refining $k_{eff} \sim k$, hence the heating coil speed chosen in our experimental procedure is adequate.

\subsection{Combined fit}

\begin{figure}[b!]
 	\centering
 	\includegraphics[width=1.\textwidth]{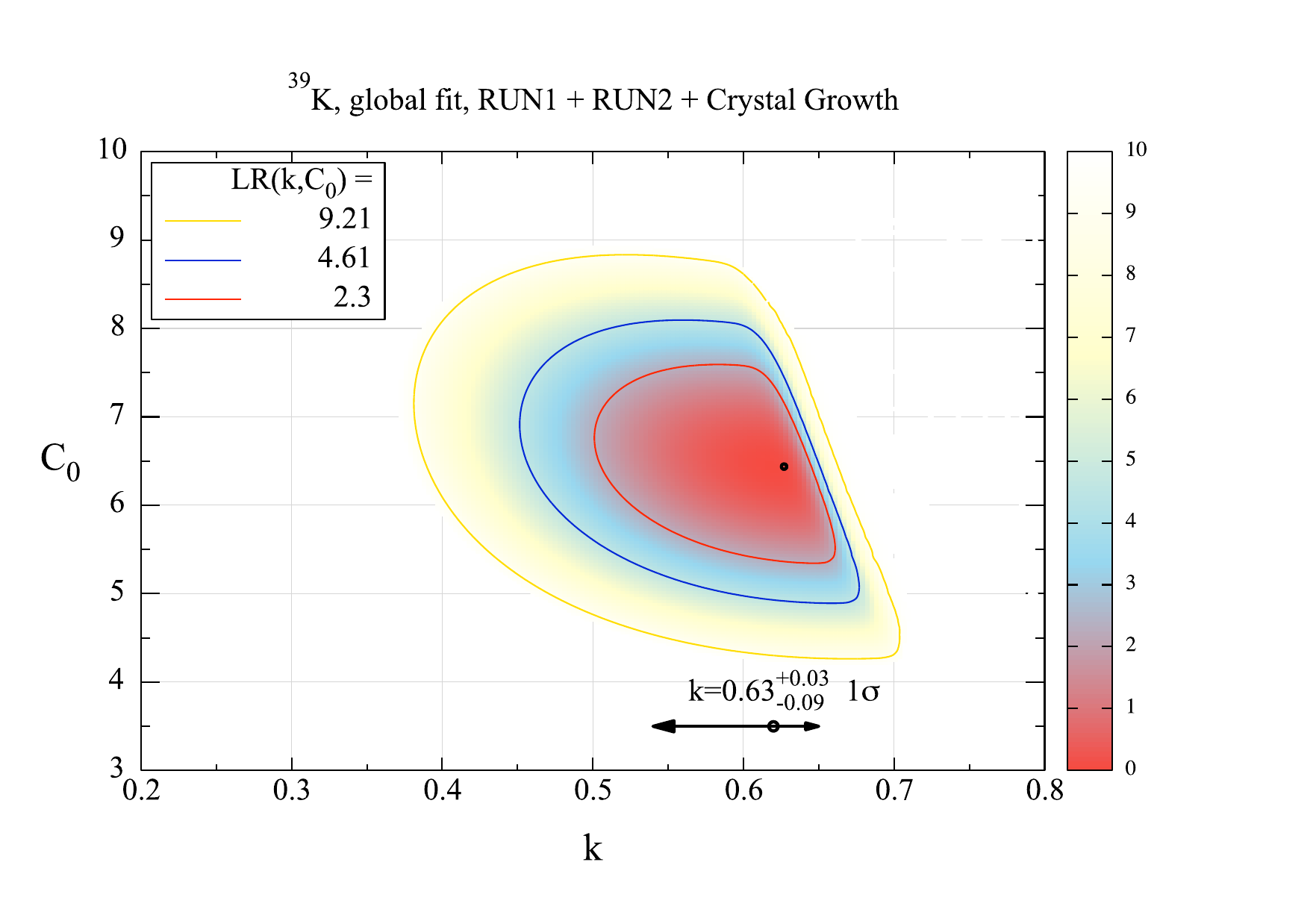}\vspace{-0.5cm}
 	\caption{Combined allowed zone in the parameter space $(C_0,k)$ for the $^{39}$K contaminant. The $1\sigma$ range for the distribution coefficient $k$ is shown. $C_0$ is in units of ppb.}
 	\label{fig:K39_Comb}
 \end{figure}

To obtain a better estimation of the distribution coefficients, we have combined the likelihoods for RUN1, RUN2, and normal freezing (tip and tail samples). For all these runs we have used the same Astro Grade powder. Therefore, the free fit parameters, namely $C_0$ and $k$, are the same for the three set of measurements. The analysis is performed simply combining the likelihood for each dataset. For illustration we show the allowed zone in the parameter space $(C_0,k)$ for $^{39}$K where all previous data are combined in Fig.~\ref{fig:K39_Comb}. 

\begin{figure}[t!]
 	\centering
 	\includegraphics[width=0.9\textwidth]{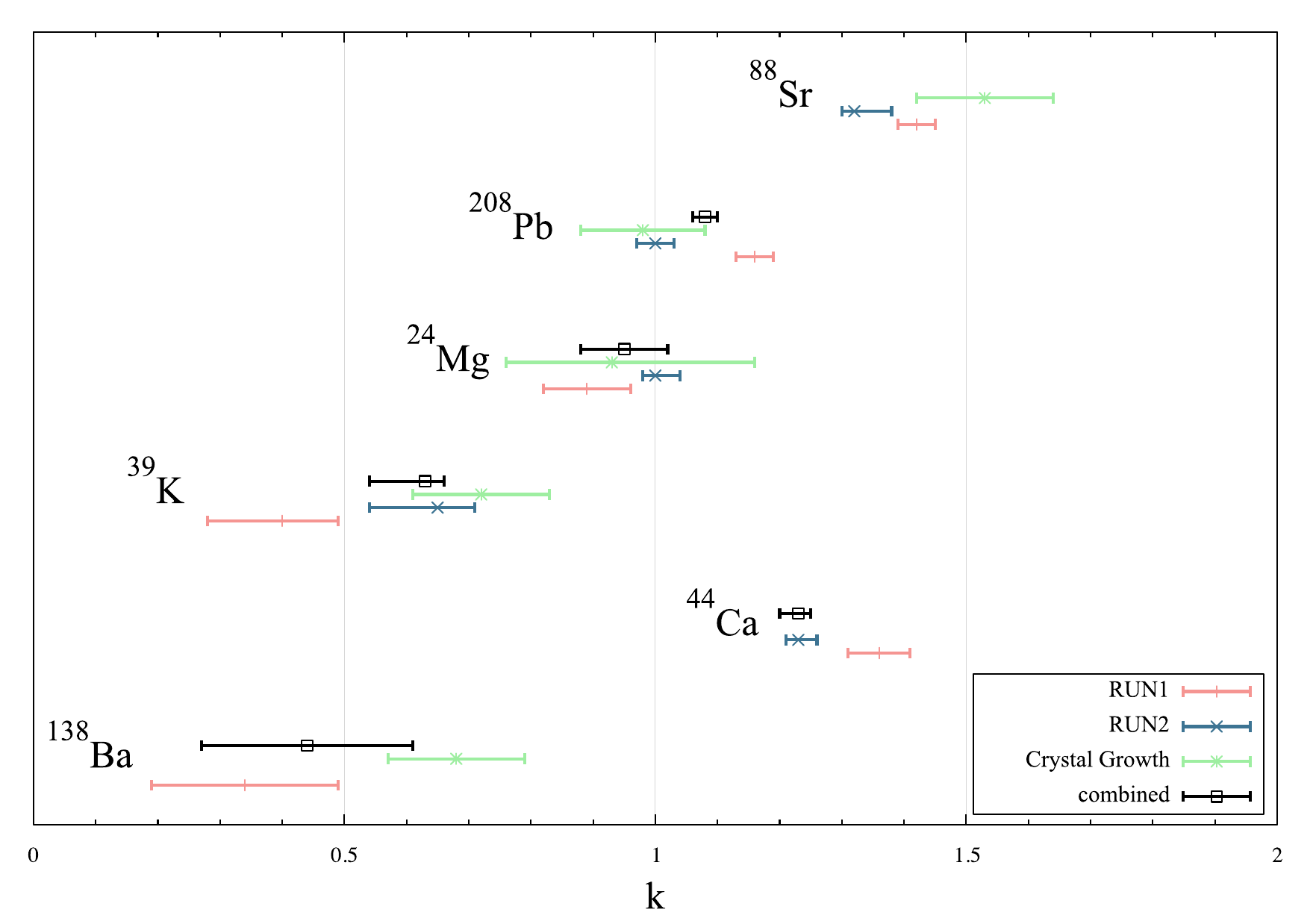}\vspace{-0.5cm}
 	\caption{$1\sigma$ ranges for the distribution coefficients $k$ for the various analysis and for the combined fit.}
 	\label{fig:range}
 \end{figure}

The combined best fit values for the distribution coefficient $k$ are summarized in Table~\ref{tab:results_comb}. For $^{138}$Ba we do not use RUN2 data for the reason explained in Sec.~\ref{sec:PR}. For $^{44}$Ca we do not use data for normal freezing since we do not have any constraint on $k$. We prefer not to combine the data for Strontium due to the strong differences for $C_0$ in any separate fit. A summary of the results for the various analysis and combined fit is shown in Fig.~\ref{fig:range}.

\begin{table}[h!]
\centering
\begin{tabular}{c|ccccccc}
\hline
& $^{138}$Ba & $^{44}$Ca    & $^{39}$K  & $^{24}$Mg  & $^{208}$Pb \\
& RUN1+CG & RUN1+RUN2    & all data  & all data & all data \\
\hline\\[-1.8ex]
$k$ & 0.44$\pm$0.17       & $1.23^{+0.02}_{-0.03}$ & $0.63^{+0.03}_{-0.09}$ & 0.95$\pm$0.07        & 1.08$\pm$0.02          \\
[1.3ex]\hline
\end{tabular}
\caption{Combined values of the distribution coefficient $k$. CG stands for Crystal Growth data.\label{tab:results_comb}}
\end{table}

\begin{table*}[t!]
\centering
\scalebox{0.9}{\begin{tabular}{c|cccccc}
\hline
 & $^{138}$Ba  & $^{44}$Ca   & $^{39}$K    & $^{24}$Mg               & $^{208}$Pb   & $^{88}$Sr (RUN2)\\
\hline\\[-1.8ex]
$f_{\rm pur}$, best
 & 0.06       & 1.21 &  0.24 & 1. & 1.1 &  1.23\\
$k\in1\sigma$ range
 & 0.03$\div$0.14 & 1.20$\div$1.22 & 0.13$\div$0.31 & 0.94$\div$1.04 & 1.08$\div$1.13 & 1.22$\div$1.24\\ \\[-1.8ex]
\hline\\[-1.8ex]
${\overline C}$, best (ppb) 
 & 0.018      & 45.1 &   1.55 & 1.97 & 1.98 & 0.61\\
 $(k,C_0)\in1\sigma$ allow.
 & 0.008$\div$0.036 & 38.6$\div$51.7 & 0.64$\div$1.80 & 1.76$\div$2.0 & 1.78$\div$2.18 &  0.61$\div$0.62\\ \\[-1.8ex]
\hline
\end{tabular}}
\caption{Coefficient of purification and final average concentration of contaminants for 80\% cut initial part of the ingot. For $^{88}$Sr we have used only the RUN2 data. \label{tab:bounds_RUN2}}
\end{table*}

\subsection{Prospect for purification}

As anticipated in Sec.~\ref{sec:examples}, after zone refining the tail is cut to obtain a purified part of the ingot (at least from contaminants with $k<1$). In particular we plan to cut and discard the 20\% final part of the purified ingot. The average remaining contamination is given by Eq.~(\ref{eq:cut}). We define the \emph{purification coefficient} the ratio between the average contamination in the remaining part of the bar and the initial contamination $f_{\rm pur}={\overline C}/C_0$.
\newline\indent
In Table~\ref{tab:bounds_RUN2} we show the values of $f_{\rm pur}$ for the six contaminants previously analyzed for the best fit value of $k$. In this table we report the $1\sigma$ range uncertainty in $k$. Moreover, since also the value of $C_0$ is uncertain, we show also the value of ${\overline C}$ for the best fit values of $k$ and $C_0$, and the maximum and minimum values varying  $(k,C_0)$ in the $1\sigma$ allowed zone in the parameter space. For $^{88}$Sr we have used only the RUN2 data.

For some contaminants (Calcium, Lead, and Strontium) the purification coefficient is $>1$. This means that the zone refining process increases the concentration of contaminants in the initial part of the ingot, although the raise is modest. For Magnesium the zone refining is practically ineffective. Comparing the powder contamination in Table~\ref{tab:measures} and the ${\overline C}$ values in Table~\ref{tab:bounds_RUN2} we note that the zone refined material has an overall reduced impurities concentration.

\section{Comparison with literature and alternative methods to determine the distribution coefficient}\label{SecVI}

We have shown how to determine the distribution coefficient from zone refining and normal freezing. In principle, phase diagrams of binary systems can also be used \cite{SuerfuPhD}. Using this method we have estimated for potassium  a distribution coefficient of $k_{\rm K} = 0.5 \pm 0.1$ derived from the KI-NaI system phase 
diagram \cite{PhaseDiagram}. In \cite{suerfu2021} it is estimated $k_{\rm K} \sim 0.57$ in NaI powder with ppb level contaminants by exploiting zone refining as in the present work. In studies on metallic impurities in NaI crystals with 100 ppm level doped contaminants a value for $k_K$ of order 0.39 is reported \cite{Gross1970}. These values are consistent with our global fit. However, we remark that in our case the contaminant concentrations are a factor $10^4$ smaller than in \cite{Gross1970}. From the same study, we can also compare the distribution coefficients for $^{44}$Ca, $^{138}$Ba, and  $^{88}$Sr to be of order 2.3, 0.4, and 1.3, respectively. 

Similarly, we analyzed the RbI-NaI system phase diagram \cite{RbINaI} to estimate $k_{\rm Rb} = 0.13 \pm 0.02$.  While current analyzed runs provided only upper limits for $^{85}$Rb, preventing a direct determination of $k_{\rm Rb}$, we plan to address this in future studies by increasing sampling near the tail of the ingot.
In \cite{Gross1970} $k_{\rm Rb}$ is estimated to be approximately 0.16 at concentrations of 100~ppm \cite{Gross1970}, suggesting a purification factor $f_{\rm pur}^{\rm Rb} \sim 10^{-3}$.

A special remark is in order for $^{208}$Pb. The PbI$_2$-NaI phase diagram \cite{PbI2-NaI} and our experimental results indicate that zone refining in NaI is ineffective for this contaminant. Reducing this specific contaminant will likely require enhancing the purity of the precursor powder prior to the zone refining stage. Recent advancements in powder purification reported by PICOLON \cite{PICOLON2025} and COSINE200 \cite{COSINE200} suggest that a combined strategy of improved chemical purification and zone refining could further enhance the final radiopurity of NaI(Tl) 
crystals.
\newline\indent
In Table~\ref{tab:results_on_k} we summarize our results on the distribution coefficients of several contaminants in NaI powder by zone refining in comparison with estimations from other methods and from the literature. We remark that differences can be due to the purity level of the powder.

\begin{table}[h!]
\centering
\renewcommand{\arraystretch}{1.3}\begin{tabular}{c|cccc}
\hline
Distribution coeff. & This work & Phase Diagram & Literature \\ \hline
\multirow{2}{1em}{$k_{\rm K}$} & $0.63^{+0.03}_{-0.09}$ & $0.5\pm0.1$ & 0.39*\cite{Gross1970} \\
         &    &              & 0.57** \cite{suerfu2021}  \\ \hline
\multirow{2}{1em}{$k_{\rm Rb}$} & - & $0.13\pm0.02$ & 0.16*\cite{Gross1970} \\
         &   &               & $<$0.59** \cite{suerfu2021} \\ \hline
$k_{\rm Ba}$ & $0.44\pm0.17$ & - & 0.4* \cite{Gross1970} \\ \hline
$k_{\rm Mg}$ & $0.95\pm0.07$ & - & 0.16* \cite{Gross1970} \\ \hline
$k_{\rm Ca}$ & $1.23^{+0.02}_{-0.03}$ & - & 2.3* \cite{Gross1970} \\ \hline 
$k_{\rm Sr}$ & $1.32^{+0.06}_{-0.02}$ & - & 1.3* \cite{Gross1970} \\  \hline
\end{tabular}
\caption{Comparison of distribution coefficients determined in this work with respect to values in the literature and estimation through the phase diagram. Numbers marked with *(**) refer to low(high) purity powder.\label{tab:results_on_k}}
\end{table}

\section{Conclusions}\label{SecVII}

This work presents the application to the large-scale production of ultra-high purity NaI(Tl) crystals for the SABRE North experiment of the zone refining method  using a custom-engineered equipment. For the first time, we have determined the distribution coefficients for several contaminants present in NaI powder at the ppb level, based on data from both commissioning and production runs. We specifically focused on potassium, which represents the dominant background source in the ROI. By combining data from normal freezing and zone refining, we obtained a distribution coefficient of $k_{\rm K} = 0.63^{+0.03}_{-0.09}$, which corresponds to about 80\% reduction of the initial contamination, when zone refining is applied according to the SABRE North experimental protocol described in this work.  

We estimate the expected background in SABRE North crystals to be approximately $0.5 \pm 0.1$~dru in the ROI for dark matter search. Note that SABRE crystals contain about 0.4\% by mass of high purity TlI to improve the light output. Our estimate assumes that there is no contamination originating from the TlI. This estimate is based on  our previous study \cite{calaprice2022} and assumes one month of surface exposure after growth when $^3$H can be produced via cosmogenic activation \cite{ANAIS-3H, Saldanha2024}. Compared to our Monte Carlo simulations for the SABRE Proof-of-Principle \cite{MCpaper}, the potassium-induced background in the ROI is projected to decrease from 0.25~dru (assuming 10~ppb contamination) to 0.06~dru. Remarkably, zone refining reduces this major background component to a level similar to that originally anticipated for the SABRE experiment when an active scintillator veto was planned. 

In conclusion, this work demonstrates the significant potential of the SABRE North crystal production strategy. By exploiting zone refining to produce ultra-high purity NaI(Tl) detectors, the experiment is well-positioned to achieve the sensitivity required for its scientific program. 

\hspace{5pt}
\section*{Acknowledgements}
The SABRE North program is supported by funding from the INFN (Italy) and the Italian Ministry of University and Research through the PRIN program and the Theoretical Astroparticle Physics (TAsP) initiative of the Istituto Nazionale di Fisica Nucleare (INFN). We acknowledge the generous support of the Laboratori Nazionali del Gran Sasso (Italy). 
We acknowledge the generous support of  Laboratorio Subterr\'{a}neo de Canfranc (Spain) for complementary ICP-MS measurements. We also thank Fausto Ortica and Aldo Romani from the Department of Chemistry, Biology, and Biotechnology at the University of Perugia for their useful discussions.

\end{document}